\documentclass[final,3p,times]{elsarticle}

\usepackage{amssymb}
\usepackage{amsmath}
\biboptions{comma,square}
\usepackage{graphicx}
\usepackage{lineno,hyperref}
\usepackage{url}
\usepackage{footnote}
\usepackage[normalem]{ulem}
\usepackage{xcolor}
\usepackage{array}
\usepackage{etoolbox}
\usepackage{todonotes}
\usepackage{soul}
\usepackage{cancel}
\usepackage[most]{tcolorbox}
\usepackage[utf8]{inputenc}
\usepackage[T1]{fontenc}
\usepackage[english]{babel}
\usepackage{xparse}

\journal{Physics Reports}

\hypersetup{
    pdfauthor={Márcio Sampaio Gomes-Filho, Luciano Calheiros Lapas, Ewa Gudowska-Nowak, Fernando Albuquerque Oliveira},
    pdftitle={The fluctuation–dissipation relations: Growth, diffusion, and beyond},
    pdfkeywords={fluctuation-dissipation, Langevin equation, stochastic processes, generalized Langevin equation},
    pdfsubject={Physics Reports},
    pdfcreator={LaTeX with hyperref},
    pdfproducer={LaTeX},
    pdfstartview={FitH},
    pdfpagemode={UseNone},
    colorlinks=true,
    linkcolor=blue,
    citecolor=blue,
    filecolor=blue,
    urlcolor=blue
}

\newcommand{\DOInote}{
  \footnotetext[0]{\textbf{Published version:} \href{https://doi.org/10.1016/j.physrep.2025.07.004}{https://doi.org/10.1016/j.physrep.2025.07.004}}
}

\begin{document}

\begin{frontmatter}




\title{The Fluctuation-Dissipation Relations: Growth, Diffusion, and Beyond\tnoteref{label1}\DOInote} 

 \author[1]{M\'arcio Sampaio Gomes-Filho} 
\affiliation[1]{organization={Center for Natural and Human Sciences, Universidade Federal do ABC},
      addressline={Av. dos Estados, 5001}, 
      city={Santo Andr\'e},
      postcode={09210-580}, 
      state={SP},
      country={Brazil}}

\author[2]{Luciano Calheiros Lapas} 
\affiliation[2]{organization={Latin American Institute of Life and Natural Sciences, Universidade Federal da Integra\c{c}\~ao Latino-Americana},
      addressline={Av. Tancredo Neves, 6731, Bloco 6, Espa\c{c}o 3, Sala 5}, 
      city={Foz do Igua\c{c}u},
      postcode={85867-970}, 
      state={PR},
      country={Brazil}}
      
\author[3]{Ewa Gudowska-Nowak} 
\affiliation[3]{organization={Mark Kac Complex Systems Research Center and Institute of Theoretical Physics, Jagiellonian University},
      addressline={prof. Stanis{\l}awa {\L}ojasiewicza 11}, 
      city={Krak{\'o}w},
      postcode={30-348}, 
      country={Poland}} 
      
\author[4]{Fernando Albuquerque Oliveira\corref{cor1}} 
\ead{faooliveira@gmail.com}

\affiliation[4]{organization={Institute of Physics, Universidade de Bras\'ilia},
      addressline={Campus Universit\'ario Darcy Ribeiro}, 
      city={Bras\'ilia},
      postcode={70910-900},
      state={DF},
      country={Brazil}}

\begin{abstract}
In this review, we scrutinize historical and modern results on the linear response of dynamical systems to external perturbations with a particular emphasis on the celebrated relationship between fluctuations and dissipation expressed by the fluctuation-dissipation theorem (FDT). The conceptual foundation of FDT originates from the definition of the equilibrium state and Onsager's regression hypothesis. Over time, the fluctuation-dissipation relation has been vividly investigated also in systems far from equilibrium, which often exhibit wild fluctuations in measured parameters. In this review, we recall the major formulations of the FDT, including those proposed by Langevin, Onsager and Kubo. We discuss the role of fluctuations in a broad class of growth and diffusion phenomena and examine the violation of the FDT resulting from a transition from Euclidean to fractal geometry. Finally, we highlight possible generalizations of the FDT formalism and discuss situations where the relation breaks down and is no longer applicable.
\end{abstract}

\begin{keyword}
 stochastic processes \sep fluctuations \sep fluctuation-dissipation theorem \sep Langevin equation \sep growth processes \sep generalized Langevin equation \sep correlation functions \sep ergodicity breaking

 \PACS 05.40.-a \sep 05.70.Ln \sep 05.40.Jc \sep 05.10.Gg \sep 81.10.-h
\end{keyword}

\end{frontmatter}

\tableofcontents


\section{Introduction} \label{sec:intro}

\begin{figure}[h]
\centering
\includegraphics[width=0.95\textwidth]{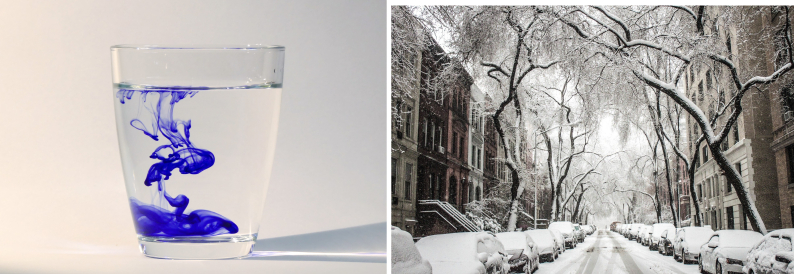}
\caption{The images from left to right illustrate the effects of fluctuations: a classical example of the diffusion of an ink drop in water~\cite{figure1} and a random snowfall on cars, streets, and trees, representing a deposition pattern that is fundamental in growth dynamics~\cite{figure2}.}
\label{fig1}
\end{figure}

Most physical observables accessible to experimental measurements consist of many components, such as particles or atoms. To understand the outcomes of these experiments and establish their predictability, it is neither possible nor reasonable to track the motion of every individual constituent. Instead, the focus shifts to analyzing deviations from the expected outcomes. Similarly, the erratic motion of pollen in a water suspension results from many collisions with solvent molecules~\cite{Brown28,Brown28a}. The jiggling (zigzag displacement) of observed pollen particles and the abrupt changes in their velocities during motion are caused by these collisions, driven by thermal noise~\cite{Langevin08,Nyquist28,Johnson28,Onsager31,Callen51,Callen52}. This thermal noise drives the motion and induces observable disturbances in trajectories. The same source of fluctuations provides energy to molecular motors~\cite{Julicher97,Astumian94,Schliwa03,Baroncini19,Chowdhury13,Astumian96,Hoffmann16,Bao03,Bao06,Nettesheim20}, powers chemical reaction pathways~\cite{Conca04}, and can amplify signals in phenomena like stochastic resonance~\cite{Gammaitoni98,Benzi81} and growth~\cite{Barabasi95,Edwards82,Kardar86}. The implications of fluctuations are critical in many areas of natural sciences, spanning from microscopic systems and molecular biology to ecology, financial market models~\cite{Gontis10}, social contact analysis~\cite{Urry04}, and climate dynamics~\cite{Chekroun11}. In Figure~\ref{fig1}, we illustrate several well-known phenomena where fluctuations play a fundamental role, such as diffusion (on the left) and growth (on the right).

The aim of this work is to review both historical and recent studies on the role of fluctuation-dissipation relations (FDRs) in describing anomalous diffusion and growth processes. While we briefly recall the foundations of linear response theory and canonical formulations of the fluctuation-dissipation theorem (FDT), our main emphasis is on modern developments that extend these ideas beyond equilibrium and into complex dynamical systems.

The review is organized as follows: We begin by briefly examining the concept of fluctuations around equilibrium. Next, we transition to the Langevin formalism and the FDT, focusing on the mechanisms by which systems are driven toward equilibrium. Section~\ref{sec:Beyond} revisits the Kubo formulation of the FDT~\cite{Kubo91} and discusses its modern applications. In Section~\ref{sec:violation}, we explore the limitations of the fluctuation-dissipation formalism and present some fundamental theorems related to it. Section~\ref{sec:growth} is dedicated to growth mechanisms, presenting the most challenging form of Langevin's growth equation~\cite{Barabasi95,Edwards82,Kardar86}, along with a discussion of the Kardar-Parisi-Zhang formalism~\cite{Kardar86}. In Section~\ref{sec:fractal}, we explore how a proper interpretation of geometry -- specifically, transitioning from Euclidean to fractal geometry -- enables the recovery of the FDT in growth. In the brief Section~\ref{sec:phase} we show how the use of a suitable fractional geometry allows us to recover the correlation function in phase transitions. Finally, Section~\ref{sec:conclusion} provides an overview of the current landscape and outlines future directions at the interface between statistical physics and non-equilibrium phenomena in growth and anomalous diffusion.

We draw the reader's attention to the fact that in order to keep authors and research area notation, each section has its own set of mathematical symbols.

\section{General concepts} \label{sec:general}

\subsection{Fluctuations around equilibrium}

Gauss's major contribution to the study of observational errors and statistical estimation in the calculation of planetary orbits~\cite{Gauss} was to assume that random measurement errors follow a normal (Gaussian) distribution, and based on this assumption, he justified the use of the method of least squares. The resulting distribution of independent deviations from the most probable value of a measured quantity $x$
\begin{equation}
P(x) = \frac{1}{\sqrt{2\pi \sigma_x^2}} \exp{\left[-\frac{\left(x-\left\langle x \right\rangle \right)^2}{2 \sigma_x^2}\right]}. \label{P1}
\end{equation}
Here, $\left\langle x \right\rangle = \frac{1}{N} \sum_{i=1}^N x_i$ represents the arithmetic mean of observed values, and $ \sigma_x^2 = \left\langle x^2 \right\rangle - \left\langle x \right\rangle ^2$ is interpreted as the precision or dispersion in a set of measurements.

Around the same time, Laplace~\cite{Laplace} successfully accounted for all the observed deviations of planets from their theoretical orbits by applying Sir Isaac Newton's theory of gravitation, developing a conceptual view of evolutionary change in the solar system's structure. Like Gauss, he showed the usefulness of a probabilistic approach for interpreting scientific data and systematized what is now known as probability theory. In this framework, the distribution in Eq.~\eqref{P1} plays a fundamental role in describing the average outcome of many statistically independent measurements and in signaling the frequency of deviations (fluctuations) around the mean.

Fluctuation phenomena are crucial in thermodynamics, particularly in nanoscale systems, where they account for a significant variability in functional properties. Thermodynamic systems at equilibrium are characterized by fluctuations that reflect the inherent discreteness of their microscopic states. The early challenges in kinetic theory, such as understanding and explaining the empirical laws of an ideal gas, were addressed through the theory of fluctuations. The probability of finding a molecule in a gas with a velocity between $v$ and $v+dv$ was associated with the distribution in Eq.~\eqref{P1}. This model worked well for a gas at high temperatures and low densities (a non-interacting ideal gas) and was confirmed experimentally. Furthermore, by considering that the total energy is purely kinetic, it was possible to determine the velocity dispersion in each direction $x$, $y$, and $z$ separately, $\sigma_{v_x^2} = \sigma_{v_y^2} = \sigma_{v_z^2} = \left\langle v_x^2 \right\rangle = \frac{RT}{M}$, where $M$ is the molecular mass of the particle, $T$ the absolute temperature, and $R$ the universal gas constant. By following this reasoning and utilizing the Boltzmann-Gibbs postulate, which establishes a relationship between the number of accessible microscopic states $W$ and the entropy $S$ of an isolated system, $ S \propto \ln W$, it became possible to theoretically determine the average energy, specific heat, and the law of ideal gases. Additionally, it was possible to prove that Eq.~\eqref{P1} is the only distribution compatible with the conservation of momentum and energy.

From the first law of thermodynamics, an infinitesimal change in the internal energy $U$ of a system can be expressed as
\begin{equation}
dU = TdS - PdV + \mu dN, \label{FL}
\end{equation}
where the thermodynamic quantities are: $T$ (temperature), $S$ (entropy), $P$ (pressure), $\mu$ (chemical potential), and $N$ (number of particles within the volume $V$). In the evolution of all quantities, it is assumed that the entropy of an isolated system reaches its maximum value at equilibrium, so fluctuations in thermodynamic variables away from this state must lead to a decrease in entropy. By considering the Boltzmann-Gibbs definition of entropy, these fluctuations can be derived by expanding deviations in entropy using a quadratic approximation and invoking the stability of the equilibrium state. Such an expansion leads to the distribution in Eq.~\eqref{P1}, centered on the mean equilibrium values of the parameters defining the system's state~\cite{Huang87,Salinas01}, with standard deviations easily determined, e.g.
\begin{equation}
\label{SigE}
\sigma_U^2 = k_B T^2 N c_v,
\end{equation}
where $\sigma_U^2$ is the standard deviation in the energy, $k_B = R/N_a$ is the Boltzmann constant, with $N_a$ as Avogadro's number, and $c_v = \frac{1}{N} \left(\frac{\partial U}{\partial T}\right)_V$ is the specific heat at constant volume. Similarly, the standard deviation in volume $\sigma_V^2$ is related to the isothermal compressibility by
\begin{equation}
\label{SigV}
\sigma_V^2 = k_B T \left\langle V \right\rangle \kappa_T,
\end{equation}
where the isothermal compressibility, $\kappa_T$, is given by:
\begin{equation}
\kappa_T = -\frac{1}{V} \left(\frac{\partial V}{\partial P} \right)_T.
\end{equation}

For magnetization $\mu$, near a phase transition, as $T$ approaches the critical temperature $T_c$ (i.e., $T \rightarrow T_c$) and the external magnetic field $h_0$ is small $h_0 \rightarrow 0$ one readily obtains~\cite{Falk69}
\begin{equation}
\sigma_\mu^2 = k_B T_c \chi_\mu,
\label{SigM}
\end{equation}
where $\chi_\mu$ denotes the magnetic susceptibility, $\chi_\mu = \partial \mu/\partial h_0$, which is the derivative of the magnetization $\mu$ with respect to $h_0$. Equation \eqref{SigM} is the standard deviation in the magnetization. Note that this relation is valid only in the linear regime, where $h_0$ is small.

A similar relation was obtained for the order parameter $r$ of a noise-induced phase synchronization in a Kuramoto model~\cite{Pinto16,Pinto17}
\begin{equation}
\label{Sigr}
\sigma_r^2 = k_B T_c \chi_r,
\end{equation}
after rephrasing the analysis of entropy in terms of the stationary probability distribution derived from the Langevin description of the system. The above relations show that the responses $c_V$, $\kappa_T$, and $\chi_\mu$ ($\chi_r$) are positive and associated with fluctuations. 

For an extensive physical system described by a state variable $x$, such that $\left\langle x \right\rangle \propto N \neq 0$, with $N$ representing the number of additive subunits contributing to $x$, the relative fluctuation of $x$ typically scales as
\begin{equation}
\label{LGN}
\frac{\sigma_x}{\left\langle x \right\rangle} \approx \frac{1}{\sqrt{N}}.
\end{equation}

This scaling arises under the assumption that the microscopic contributions to $x$ are independent and identically distributed with finite variance. As such, it breaks down in situations where these assumptions no longer hold, particularly near critical points, where long-range correlations emerge and large collective fluctuations dominate.

For example, close to a phase transition, thermodynamic quantities such as the volume or energy may exhibit anomalously large fluctuations, violating the scaling in Eq.~\eqref{LGN}. A paradigmatic case is the vicinity of the solid-liquid transition, where the volume of a sample of ice can fluctuate by more than 10\% of its mean value. Such behavior is captured by the divergence of $c_V$ and $\kappa_T$, as seen in Eqs.~\eqref{SigE} and \eqref{SigV}. Some mean-field theories, such as the Van der Waals equation, exhibit an unstable region where $\kappa_T < 0$. These macroscopic signatures of criticality highlight the fundamental role of fluctuations in statistical mechanics, which, although occasionally compatible with Gaussian statistics, often necessitate going beyond such approximations. For further details on the statistical origin of fluctuations and the breakdown of Gaussian behavior near phase transitions, we refer the reader to standard textbooks such as Refs.~\cite{Huang87,Salinas01,Paul13}, as well as to recent expositions on the role of large deviations and correlated fluctuations~\cite{Vulpiani2014}.

\subsection{Fluctuations out of equilibrium} \label{subsec:fluctuations_out_of_equilibrium}

When analyzing systems out of equilibrium, we can observe that they are more complex, but they also present a richness of new phenomena, such as observable diffusion processes. The diffusion is the simplest and most general fluctuation mechanism in out of equilibrium systems. This dynamical mechanism transports particles, energy, or even information toward equilibrium~\cite{Morgado02,Costa03,Metzler99,Sancho04,Lapas07,Lapas08, Weron10,Thiel13,Dorea06,Mckinley18,Flekkoy17}. This transport occurs when the system is not homogeneous. The inhomogeneous medium creates gradients that drive the system towards equilibrium.

For two centuries~\cite{Brown28,Brown28a,Vainstein06,Santos19, Metzler00,Metzler04}, the diffusion mechanism has been intensively investigated because it plays a crucial role in many types of systems. The well-known experiments performed by Robert Brown drew attention to the stochastic trajectories of microscopic pollen grains~\cite{Brown28}, which he initially associated with life. Later, he observed the same phenomenon in coal dust~\cite{Brown28a}. In this way, he discovered a new and intriguing phenomenon, now known as Brownian motion. Although the individual trajectories seem indescribable, their averages provide important information. For example, if we track the particle's position as a function of time along the $x$-axis, we find the average displacement $\left\langle x \right\rangle = 0$ and its mean square
\begin{equation}
\label{x20}
\sigma_x^2 = \left\langle \left(\Delta x(t)\right)^2 \right\rangle,
\end{equation}
where $\Delta x = x(t) - x(0)$ is the particle's displacement from the origin in time $t$. Although often referred to as part of Einstein's theory of Brownian motion, the equation
\begin{equation}
\label{eq.sqr_disp}
\sigma_x^2 = 2Dt,
\end{equation}
expresses the MSD for normal diffusion in one dimension. It is a key result that emerges from the diffusion equation and was indeed consistent with the theoretical framework introduced by Einstein in 1905. The relation commonly known as the Einstein relation establishes a connection between the diffusion coefficient \( D \) and the particle's mobility \( \varsigma \) through the thermal energy~\cite{Einstein1905,Oliveira19}
\begin{equation}
  D=\varsigma k_B T.
  \label{eq.Eintein.rel}
\end{equation}
Here $\varsigma$ refers to the ratio of the drift velocity over applied force and measures how easily the particle moves in response to a force. In a viscous medium the drag force on a particle is proportional to the drift velocity multiplied by a friction coefficient $\gamma$, hence $\varsigma=({m} \gamma)^{-1}$.

In the next section, we discuss Langevin's work on diffusion, where he introduced the first stochastic equation of motion for a particle and the first form of the FDT. As we shall see later, we can write its generalization as Mori's equation and the Kubo FDT.

\subsection{The Langevin equation and the fluctuation-dissipation theorem}\label{sec:lange}

In the early 20th century, atomic theory was still highly controversial. For some researchers, the unanswered questions were enormous problems, while for others, they were a great stimulus. The works of Einstein and Smoluchowski on Brownian motion had a significant impact. Under different approaches, P. Langevin, A.D. Fokker, M. Planck, and A. Kolmogorov studied Brownian motion as a special class of stochastic Markov processes~\cite{Langevin08,Vainstein06,Oliveira19, Einstein1905,Einstein56,Nowak17}.

Langevin initiated the era of explicit stochastic processes in physics by considering the equation of motion for a particle moving in a fluid~\cite{Langevin08}
\begin{equation}
m\frac{dv(t)}{dt} = -m\gamma v(t) + f(t),
\label{L}
\end{equation}
where $m$ and $\gamma$ are the mass of the particle and the friction coefficient, respectively. Langevin's ingenious and elegant proposal was to model the complex interactions between particles by considering two dominant forces. The first represents a frictional force, $-m\gamma v$, where the characteristic time scale is $\tau = \gamma^{-1}$, while the second is a stochastic force, $f(t)$, with a time scale $\Delta t \ll \tau$, related to the random collisions between the particle and the fluid molecules.

The fluctuating force $f(t)$ in Eq.~\eqref{L} obeys the following conditions:
\begin{itemize}
\item[$(i)$] The mean force due to the random collisions on the particle is zero:
\begin{equation}
\label{fmed}
\langle f(t)\rangle = 0.
\end{equation}
\item[$(ii)$] There is no correlation between the initial particle velocity and the random force:
\begin{equation}
\label{fv}
\langle f(t)v(0)\rangle = 0.
\end{equation}
\item[$(iii)$] The fluctuating force at different times $t$ and $t'$ is uncorrelated:
\begin{equation}
\label{whitenoise1}
\langle f(t)f(t')\rangle = B \delta(t-t'),
\end{equation}
\end{itemize}
where $B$ is a constant to be determined. To understand the dynamical properties of a particle that obeys the equation of motion~\eqref{L} and conditions~\eqref{fmed} to~\eqref{whitenoise1}, we start with the formal solution:
\begin{equation}
\label{v0}
v(t) = v(0)\exp{(-\gamma t)} + \frac{1}{m}\int_0^t \exp[-\gamma(t-t')]f(t')dt'.
\end{equation}
With Eq.(\ref{v0}) we can calculate the averaged squared velocity:
\begin{eqnarray}
\langle v(t)^2\rangle&=& v(0)^2\exp(-2\gamma t)+\frac{e^{-2\gamma t}}{m^2}\int^t_0\int^t_0 e^{\gamma(t'+s)}\langle f(t')f(s)\rangle dt'ds \nonumber\\
&=& v(0)^2\exp{(-2\gamma t)}+\frac{B}{m^2}\int^t_0dt'\exp{(2\gamma t')},  
\end{eqnarray}
where we have used properties of the stochastic force $f(t)$.
After a long time, the system reaches equilibrium, which means that
\begin{equation}
\label{eqt}
\lim_{t\rightarrow\infty}\langle v^2(t) \rangle = \langle v^2 \rangle_{eq} = \frac{k_B T}{m},
\end{equation}
where the subscript ``$eq$'' denotes the equilibrium ensemble average, and the expression follows from the equipartition theorem. We then use the above conditions to obtain $B = 2m \gamma k_B T$, thus relating the noise amplitude to the friction coefficient in the first explicit form of FDT:
\begin{equation}
\langle f(t)f(t') \rangle = 2m \gamma k_B T \delta(t-t').
\label{FDT0}
\end{equation}
The Langevin approach is based on replacing the effect of many frequent collisions with fluid molecules by an effective stochastic external force $f(t)$. Owing to the random character and the statistical independence of large number of individual ``kicks'' contributing to $f(t)$, the overall random force in Eq.~\eqref{L} will tend to follow a Gaussian distribution in the limit, regardless of the original distribution of each individual contribution. In this way a many-body problem becomes reduced to a much more tractable effective single-body problem with a white ($\delta$-correlated), Gaussian noise $f(t)$.

Note that Eqs.~\eqref{SigE}-\eqref{Sigr} establish a relationship between fluctuations and static (equilibrium) response of the extensive property under variation of the conjugated intensive variable (e.g. a magnetic field), that can be easily measured. In turn, by adapting Langevin's approach we deal with a dynamic equation from which we can compute various response functions and treat systems that are not in equilibrium, but close to it. Let us conclude analysis of the diffusion process Eq.(\ref{L}) by calculating the MSD. We use a trick used by Chandrasekhar~\cite{Chandrasekhar43} and start with multiplying both sides of Eq.(\ref{L}) by $x$ and taking average with respect to the fluctuating force:
\begin{equation}
\left\langle x\frac{d^2x}{dt^2} \right\rangle + \gamma\left\langle x\frac{dx}{dt}\right\rangle 
=\frac{1}{2}\frac{d}{dt}\left \langle\frac{dx^2}{dt}\right\rangle-\left\langle v^2\right\rangle+\frac{\gamma}{2}\left\langle\frac{dx^2}{dt}\right\rangle=0.
\end{equation}
Assuming stationarity (taking limit $t\rightarrow\infty$) and using formula Eq.(\ref{eqt}) we arrive at 
\begin{equation}
  \left\langle dx^2\right\rangle =\frac{2k_BT}{\gamma m}dt,
\end{equation}
or otherwise $\lim_{t\rightarrow\infty}\frac{\langle x^2(t)\rangle}{t}=2D$, where $D$ is the diffusion constant of the Brownian particle. We can obtain a useful relation between $D$ and velocity autocorrelation function by observing that
\begin{equation}
\label{eq.sqr2}
\left\langle \left( \Delta x(t) \right)^2 \right\rangle = \int_0^t dt^{\prime} \int_0^{t} dt^{\prime\prime} \langle v(t^{\prime})v(t^{\prime \prime}) \rangle,
\end{equation}
where the integrand stands for
\begin{equation}
\label{Cv0}
C_v(t) = \left\langle v(t+t')v(t') \right\rangle\equiv\left\langle v(t)v(0)
\right\rangle.
\end{equation}
The equivalence of the second term in Eq.(\ref{Cv0}) follows from stationarity (thermalization of velocities), where the expectation value of any observable is invariant under time translation. By taking time derivative of $C_v(t)$ and making use of Eq.~\eqref{L} we obtain
\begin{equation}
\label{Cv1}
\frac{d}{dt} C_v(t) = \left\langle \dot{v}(t)v(0) \right\rangle=-\gamma C_v(t),
\end{equation}
which can be solved analytically, yielding the solution
\begin{equation}
\label{Cv2}
C_v(t) = \frac{k_B T}{m}\exp(-\gamma t).
\end{equation}
Thus, by introducing Eq.~\eqref{Cv2} into Eq.~\eqref{eq.sqr2}, we obtain an equation similar to Eq.~\eqref{eq.sqr_disp}, with the diffusion coefficient given by
\begin{equation}
\label{Dif}
D = \frac{k_B T}{m \gamma},
\end{equation}
in line with the relation obtained by Einstein, Eq. \eqref{eq.Eintein.rel}. It is worth noting that this result can also be expressed as an integral of the velocity autocorrelation function:
\begin{equation}
\label{Kubo}
D = \lim_{t\rightarrow\infty}\int_0^t\langle v(t')v(0)\rangle dt'=\int_0^{\infty}C_v(t)dt.
\end{equation}
This expression is sometimes called the Green-Kubo formula \cite{Green1954,Kubo1957}. It indicates that diffusion coefficient of the Brownian particle can be measured by estimation of the long-time (equilibrium) behavior of its velocity autocorrelation function. Although we have performed our analysis for one-dimensional motion, the generalization of Eq.~\eqref{L} to $d$ dimensions can be achieved by simply multiplying the result of Eq.~\eqref{eq.sqr_disp} by $d$~\cite{Einstein1905,Einstein56}.

It is important to note that, following the theoretical work of Einstein, Langevin, and Smoluchowski, numerous experimental results have confirmed the theory of Brownian motion.

\subsubsection{The fluctuation-dissipation mechanism driving to equilibrium}

As discussed above, Langevin's mechanism drives the system toward equilibrium. In this subsection, we demonstrate this result for a complex system and provide a more detailed discussion of this mechanism.

\begin{figure}[h]
\centering
\includegraphics[width=0.95\linewidth]{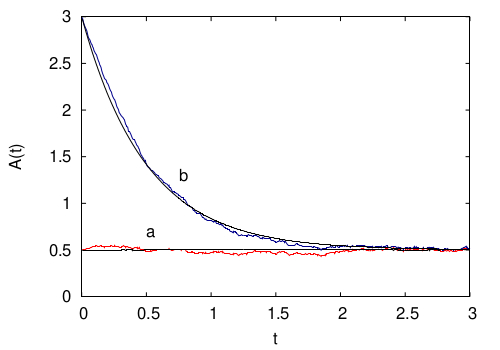}
\caption{The evolution of the instantaneous average kinetic energy $A(t)$ as a function of {time $t$} for various temperatures: (a) $T_0 = T$; (b) {$T_0 = 6T$}. Adapted from reference~\cite{Vainstein06} with permission from Springer Lecture Notes in Physics.}
\label{fig_Ene}
\end{figure}

First, let us solve Eq.~\eqref{L} for a system of $N$ particles. Using $\tau = \gamma^{-1}$ and $\Delta t = 0.005 \tau$, the Brownian force $f(t)$ is taken to be constant during each time increment and of magnitude $r \sqrt{6m\gamma k_B T/\Delta t}$, with random numbers $r$ uniformly distributed in the interval $-1 \leq r \leq 1$ and $T$ representing the reservoir temperature. The initial conditions for the numerical integration were set with uncorrelated non-Gaussian velocity distributions in the form $v(0) = \pm \sqrt{k_B T_0/m}$. Figure~\ref{fig_Ene} displays the instantaneous average kinetic energy per particle, $A(t) = \left\langle \frac{1}{2}mv^2(t) \right\rangle$, as a function of time for a typical simulation~\cite{Vainstein06}. Energies are given in units of $k_B T$ and time in units of $\tau$. The floating lines are plotted for $N = 500$ particles and show large fluctuations, while the continuous lines represent an ensemble of $N = 10^6$ particles. On this scale, there is no visible difference between the exact results in $\left\langle v^2 \right\rangle$ and the simulations for $N = 10^6$ particles. By setting the initial temperature $T_0$ for each independent system, (a) $T_0 = T$ and (b) {$T_0 = 6T$}, one observes both curves plateau after reaching thermal equilibrium. Thus, the fluctuations are consistent with those expressed by Eqs.~\eqref{SigE} and~\eqref{LGN}.

In the example presented above, neither the initial conditions for the velocities nor the noise followed a Gaussian distribution. While this may raise concerns, it is not inherently problematic. Ultimately, the resulting distribution of velocities converges to a Gaussian form. The remaining issue pertains to the nature of the noise, which we address below.

For a system of interacting particles, we need only add the interaction force into Eq.~\eqref{L}. Consider, for example, a classical chain with $N$ monomers~\cite{Oliveira94,Oliveira95,Oliveira00}. The Langevin equation for a particle of mass $m$ at site $i$, with $i=1,2,\cdots, N$, is given by
\begin{equation}
\label{chain1}
m\frac{d^2q_i}{dt^2} = -m\gamma \frac{dq_i}{dt} + F(a+q_i-q_{i-1}) - F(a+q_{i+1}-q_i) + f_i(t),
\end{equation}
where $F$ is the force between neighboring particles, $a$ is the equilibrium position (where $F(a) = 0$), $q_i$ is the displacement from the equilibrium position, and $f_i(t)$ is a random force.

A straightforward method to renormalize the equation of motion is to perform a decimation~\cite{Oliveira95,Oliveira00}, which involves rewriting the equations of motion to eliminate intermediate sites. For example, if the interaction is between a site $i$ and its neighbors $i \pm 1$, we eliminate these sites to get an interaction between site $i$ and its next-nearest neighbors $i \pm 2$. Let us define $y_i = q_{i+1} - q_i$ and subtract two successive equations from Eq.~\eqref{chain1} to obtain
\begin{equation}
\label{chain2}
m\frac{d^2y_i}{dt^2} = -m\gamma \frac{dy_i}{dt} - F(a-y_{i-1}) + 2F(a+y_i) - F(a-y_{i+1}) + [f_{i+1} - f_i(t)].
\end{equation}

Note that renormalization for the positions is straightforward if $F$ is linear~\cite{Oliveira00}, $F(a+y) = -Cy$, with $C$ as a constant; for nonlinear forces, it is much more complicated~\cite{Oliveira95}. However, for noise that is independent of the lattice, renormalization is trivial. For simplicity, let us consider the noise as discrete $f_i = \Lambda j$, with $j = \pm 1/2$, intensity $\Lambda$, and probability $W_0(j) = 1/2$. In Eq.~\eqref{chain2}, $f_{i+1} - f_i$, or equivalently $f_{i+1} + f_i$, now takes values $j = 0, \pm 1$ with probabilities $W_1(0) = 1/2$ and $W_1(\pm 1) = 1/4$. For a general iteration of order $n$, we have
\begin{equation}
\label{Wn}
W_n(j) = \sum_{l} W_{n-1}(l)W_{n-1}(j-l),
\end{equation}
i.e., the sum of all $l$ and $j-l$ that gives a $j$. $W_n(j)$ is a binomial distribution~\cite{Oliveira00}, similar to that found in the study of unbiased random walks~\cite{Salinas01}. However, it is intriguing to derive it from the equations of motion. As $n \rightarrow \infty$, $W_n(j)$ becomes Gaussian, as required by the central limit theorem (CLT) and alternatively by the probabilistic interpretation of renormalization group (RG) transformations~\cite{Jona01}.

\begin{figure}[!h]
\centering
\includegraphics[width=0.95\linewidth]{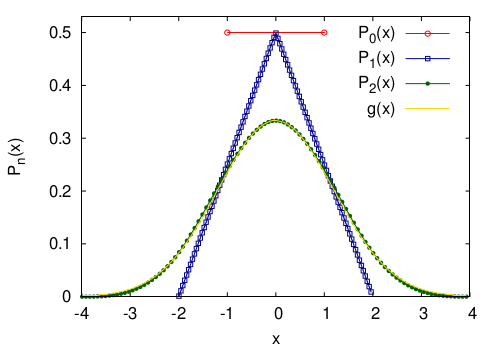}
\caption{The probability density $P_n(x)$ of obtaining a noise with intensity between $x$ and $x+dx$ as a function of $x$. The evolution of $P_n(x)$ is shown for $n = 0, 1$ and $2$, starting from the uniform distribution $P_0(x)=1/2$. The Gaussian function $g(x)= \exp{-(\pi x^2/9)}/3$ is also plotted, scaled to match the peak height of $P_2(x)$. From reference~\cite{Oliveira95} with permission from PRB.}
\label{fig_Px}
\end{figure}

If we consider the noise distribution as continuous from the very beginning, after $n$ iterations, we get the convolution
\begin{equation}
\label{convolution}
P_{n}(x) = \int_{-L_{n-1}}^{L_{n-1}}P_{n-1}(y)P_{n-1}(x-y)dy,
\end{equation}
where $L_n = 2L_{n-1}$. For example, if we consider random numbers $x$ uniformly distributed in the interval $-1 \leq x \leq 1$, i.e., $P_0(x) = 1/2$, and $L_0 = 1$, we obtain from the convolution~\eqref{convolution}
\begin{equation}
\label{P11}
P_1(x) = 
\begin{cases}
\frac{1}{4} \left(2-|x| \right), &\text{ for } |x| \leq 2, \\
0, &\text{ for } |x| > 2,
\end{cases}
\end{equation}
and by using Eq.~\eqref{convolution} again, we obtain
\begin{equation}
\label{P2}
P_2(x) = 
\begin{cases}
\left(32-12x^2+3|x|^3 \right)/96, &\text{ for } |x| \leq 2, \\
\left(4-|x| \right)^3/96, &\text{ for } 2 \leq |x| \leq 4, \\
0, &\text{ otherwise. }
\end{cases}
\end{equation}
In Figure~\ref{fig_Px}, we plot the probability density $P_n(x)$ as a function of $x$ for $n=0,1,2$. Note that $P_2(x)$ is almost Gaussian.

It is also interesting to note that if we apply the Fourier transform to Eq.~\eqref{convolution}, we get
\begin{equation}
\label{Pnk}
\tilde{P}_n(k) = \tilde{P}_{n-1}(k)^2,
\end{equation}
which we can denote as the invariants of the renormalization group transformation (IRGT), the functions that retain their form under the transformations.

A possible set of IRGTs is given by $\tilde{P}_n(k) = \exp(-(c_n k)^\nu)$, where the constant $c_n$ follows the scaling $c_{n+1} = 2^{1/\nu}c_n$. As $n \rightarrow \infty$, the IRGT with $0 \leq \nu \leq 2$ yields the L\'{e}vy functions~\cite{Levy37}
\begin{equation}
\label{Levy}
L(\nu,x) = \frac{1}{2\pi} \int e^{ikx} e^{-(c|k|)^\nu}dk.
\end{equation}

For $\nu=1$, we obtain a Lorentzian, and for $\nu=2$, we obtain a Gaussian curve. Thus, Eq. \eqref{convolution} as $n \rightarrow \infty$ and $P_n(x) = P(x)$ leads us to the generalized central limit theorem (GCLT), which has also been discussed in the context of RG fixed points and the emergence of universality classes for sums of strongly correlated variables~\cite{Jona01}. Note that if we study, for example, the kinetic energy as in Figure~\ref{fig_Ene}, we obtain $A(t \rightarrow \infty) = \frac{1}{2}k_B T$, i.e., a finite moment. Since in the set of functions~\eqref{Levy}, Gaussians are the only distributions with all finite moments, Gaussian distributions are the most expected outcome after equilibrium is reached. Note that Eq.~\eqref{Wn} or Eq.~\eqref{convolution} is independent of the forces $F$ between neighbors, making them very general. The extraordinary success of Langevin's approach to nonequilibrium statistical mechanics is what we might call Langevin's legacy, as we shall further explore.

The elegant and straightforward Langevin approach aids in addressing many situations in physics. It allows for analytical calculations in simplified models and even provides limiting results for more complex systems. Additionally, it is easy to perform computer simulations within his framework. Therefore, the Langevin equation and its quantum formulation, with the concept of fluctuating fields, have opened a vast area of investigation across many systems, such as light scattering~\cite{Leite64,Penna76,Marcus60,Oliveira81,Loudon00,Santos00,Benmouna01}, neutron scattering in liquid metals~\cite{Rahman62,Yulmetyev03}, anomalous statistics of laser-cooled atoms~\cite{Afek23}, polymer chain dynamics~\cite{Florencio85,Odell94,Toussaint04,Doerr94,Oliveira96,Oliveira98a,Gonzalez99,Maroja01,Dias05,Sain06,Azevedo16}, molecular motors~\cite{Bao06,Bao03a,Qiu19}, conductivity~\cite{Dyre00,Oliveira05}, reaction rate theory~\cite{Kramers40,Oliveira98,Hanggi77,Hanggi90}, catalytic reactions~\cite{Evans90,Evans91,santos07,Santos24} and noise synchronization~\cite{Boccaletti02,Longa96,Maritan94,Ciesla01,Osipov07,Morgado07}. 
The application of diffusion concepts finds many uses in science, for example, in controlled drug delivery, where understanding the drug release mechanisms and characteristic release times is essential~\cite{Siepmann12,GomesFilho16,Ignacio17,Singh19,mircioiu2019,GomesFilho20,Yang20,GomesFilho22,Vigata20,Salahshoori24}. However, all these phenomena are just a small part of the vast field of anomalous diffusion~\cite{Astumian94,Bao03,Morgado02,Sancho04,Lapas08,Weron10,Thiel13,Dorea06,Mckinley18,Santos19,Vainstein08,Donato06}.

\subsection{Linear regime and fluctuation-dissipation relation}

Let us briefly recapitulate approach to dissipation of energy under perturbations as proposed by Callen~\cite{Callen51} and Kubo~\cite{Kubo1957}. 
Both frameworks show that equilibrium time correlations of an observable $A$ which couples the system to an external forcing $h(t)$ determine how a system absorbs and irreversibly disperses energy when influenced by an external force. The total Hamiltonian takes the form $H_{tot}=H-Ah(t)$. For small perturbation, linear response requires
\begin{equation}
  \left\langle A(t) \right\rangle =\int dt' \chi(t,t')h(t'),
\end{equation}
or in a frequency domain $\left\langle A(\omega) \right\rangle =\chi(\omega)h(\omega)$, where $\chi$ stands for the response function assumed invariant under time translations. The time evolution of $\left\langle A(t) \right\rangle$ can be otherwise expressed by
\begin{equation}
 \left\langle A(t) \right\rangle = Z_{tot}^{-1}\int_VdV\exp{(-\beta H_{tot})}A(t), 
\end{equation}
where $Z_{tot}$ is a normalization factor. In the linear approximation (first order with respect to perturbation field) one obtains
\begin{equation}
  \left\langle A(t) \right\rangle -\left\langle A(t) \right\rangle_{eq}=\beta h\left (\left\langle A(0)A(t) \right\rangle_{eq} -\left\langle A(0) \right\rangle_{eq} \left\langle A(t) \right\rangle_{eq}\right )
  \label{eq:<A>}
\end{equation}
with subscript ``$eq$'' referring to equilibrium canonical ensemble with unperturbed Hamiltonian. From Eq.~\eqref{eq:<A>} one gets 
\begin{equation}
\chi(t)=-\beta \Theta(t)\frac{d}{dt}\left\langle A(0) A(t) \right\rangle_{eq},
\end{equation}
or in the Fourier domain $\mathrm{Im}~\chi(\omega)=-\beta \frac{\omega}{2}C_A(\omega)$, where $C_A(t)=\left\langle A(0)A(t) \right\rangle_{eq}$
This linear response framework can be extended to quantum systems with few adjustments: interpreting averages as traces over the quantum density matrix and replacing the time derivative of the correlation function of classical variables by the commutator of the quantum observables at different times.

A comprehensive overview of contemporary nonequilibrium statistical physics covering Langevin equations, general stochastic processes, kinetic theory, fluctuation theory, linear response theory, and stochastic thermodynamics with various applications, including climate systems, information theory, granular and active matter has been recently presented in a series of books~\cite{Gaspard,Politi,Sarracino25}.

\section{Fluctuation theorems beyond Langevin framework} \label{sec:Beyond} 

Langevin's versatile framework has proven to be an exceptionally powerful tool for describing stochastic dynamics in physical systems. It can be extended in two important ways:
\begin{enumerate}
  \item To incorporate memory effects and non-Markovian dynamics, including temporally correlated noise (i.e., colored noise), valid for both classical and quantum systems.
  \item To incorporate quantum effects, including quantum fluctuations and dissipation, by modeling the system's interaction with a microscopic environment-typically represented as a heat bath of harmonic oscillators.
\end{enumerate}

These topics have been the focus of extensive investigation since the foundational work in statistical and quantum mechanics, and they remain at the forefront of ongoing research efforts.

To extend the classical theory of the previous section into a broader setting, we consider a system interacting with a environment (or heat bath) modeled as a collection of free harmonic oscillators \cite{Ford,Ford1,Caldeira83}. Through the application of fundamental physical principles, such as causality and the uniqueness of the equilibrium state, a macroscopic Langevin-like equation for a reduced description of the system can be derived. It is important to note, however, that this approach, often termed ``phenomenological'' by Van Kampen~\cite{vanKampen92,vanKampen97}, is neither unique nor straightforward. In quantum mechanics, the physical origins of both fluctuations and damping must be carefully considered~\cite{vanKampen97}.

Over the past century, significant advances have been made in the linear response theory of quantum systems to external perturbations, inspired by the pioneering works of Nyquist, Johnson, Kubo, and Callen~\cite{Nyquist28,Johnson28,Callen51,Callen52,Takahashi52,Kubo1957,Kubo57a}. Notably, it has been established that the linear response coefficients in Hamiltonian systems are proportional to two-point correlation functions \cite{Kubo1957,Kubo57a}, similar to those encountered in stochastic systems, as discussed in Subsection \ref{subsec:fluctuations_out_of_equilibrium}.

Further extensions and connections to thermodynamics have been developed, in particular by Onsager~\cite{Onsager31}, who investigated the relaxation dynamics of macroscopic observables in terms of the first time derivative of entropy, which is nonzero away from equilibrium. According to Onsager's formulation, the derivative of entropy with respect to a macroscopic variable is termed the thermodynamic force, which, in this context, is linearly coupled to a thermodynamic flux by a kinetic coefficient. In his seminal work, Onsager also provided a proof of symmetry relations in kinetic coefficients and established a connection between the dispersion of fluctuations and relaxation rates. It took several decades for further developments to emerge, particularly in the context of stochastic thermodynamics of mesoscopic systems, the linear response of non-Markovian processes, and the response of stochastic processes driven by non-Gaussian fluctuations~\cite{Kubo91,Costa03,Hanggi90,Deker75,Kubo74,Ruelle98,Zwanzig01,Seifert12,Dybiec12,Dybiec17,Nowak14,Marconi08,Sarracino19,Winkler20}. Another significant generalization of FDRs stems from the pioneering work of Bochkov and Kuzovlev~\cite{Bochkov77,Bochkov81}, who derived rigorous equalities for fluctuations (fluctuation theorems) in arbitrary systems subjected to strong nonequilibrium conditions. Their work was later expanded by Jarzynski~\cite{Jarzynski97,Jarzynski00} and Crooks~\cite{Crooks99}. Notably, in contrast to the FDR studied in the linear regime, fluctuation theorems provide insights into nonlinear response and have been experimentally verified~\cite{Saira12}.

\subsection{Mori's equation and the Kubo fluctuation-dissipation theorem}

Mori~\cite{Kubo74,Zwanzig01,Mori65,Mori65a} developed a method that generalizes the Langevin equation into the form
\begin{equation}
\label{GLE}
\frac{d A(t)}{d t}=-\int _{0}^{t}\Gamma (t-t')A(t')\,d t'+F(t),
\end{equation}
where $F(t)$ is a stochastic noise subject to the conditions
\begin{equation}
\label{F0}
\langle F(t) \rangle =0,
\end{equation}
\begin{equation}
\label{F1}
\langle F(t)A(0) \rangle =0,
\end{equation}
and
\begin{equation}
\label{fdt}
C_F(t-t')=\langle F(t)F(t') \rangle = \langle A^2 \rangle_{eq} \Gamma(t-t').
\end{equation}

Equation~\eqref{GLE} is known as the generalized Langevin equation (GLE), also referred to as Mori's equation. It provides a unifying formalism valid for both classical and quantum systems. In the classical case, $A(t)$ is a phase-space variable and all averages $\langle \cdot \rangle$ denote ensemble averages over an appropriate probability distribution. In the quantum case, $A(t)$ represents a Heisenberg-picture operator and $\langle \cdot \rangle$ denotes a statistical average over a density matrix (i.e., a quantum trace). Both Eqs.~\eqref{F0} and~\eqref{F1} resemble the Langevin equation (Eqs.~\eqref{fmed} and~\eqref{fv}), but the third condition, Eq.~\eqref{fdt}, reveals a more general behavior. The presence of the memory kernel $\Gamma(t)$ signifies that the system's dynamics at each instant depend on its history, implying that the fluctuating force $F(t)$, while random, exhibits temporal correlations. This memory effect characterizes the process as non-Markovian. In this framework, Eq.~\eqref{fdt} embodies the Kubo FDT~\cite{Kubo91,Kubo66}. The combination of Mori’s projection operator formalism and the Kubo FDT leads naturally to the GLE, Eq.~\eqref{GLE}.

An alternative yet related approach was developed by Lee~\cite{Lee83,Lee82,Lee83a,Lee84}, who proposed a method of recurrence relations to derive time correlation functions, bypassing the projection operator technique. This method has been successfully applied to a wide range of quantum systems, including dense electron gases, transverse Ising model, Heisenberg model, XY models, Heisenberg models with Dzyaloshinskii-Moriya interactions~\cite{Florencio20}. It has also been used in the stochastic analysis of quantum random-exchange Heisenberg chains~\cite{Vainstein05} and in classical systems, such as harmonic oscillator chains~\cite{Florencio85}.

The derivation and interpretation of time correlation functions remain a central topic in modern statistical physics~\cite{Nowak22,Florencio20,Chen20,Feng20,Reggiani20,Maes20,Fotso20,Wu20}. Mori’s formalism generalizes the Langevin description in two important ways: it naturally accommodates quantum systems, and it incorporates correlated (colored) noise. Nonetheless, the GLE and its associated methods can also be consistently applied to classical systems, making it a versatile tool for studying dissipative dynamics across disciplines.

For instance, consider the specific case where $A(t) = mv(t)$ and
\begin{equation}
 \Gamma(t)= 2 \gamma \delta(t).
\end{equation}
In this case, the generalized Eq.~\eqref{GLE} reduces to the standard Langevin equation~\eqref{L}.

\subsubsection{Response functions}

As we have seen in the Langevin framework, the correlation function, Eq.~\eqref{Cv1}, is fundamental in determining the dynamics. For the Mori's equation, the correlation function has been a subject of intense investigation~\cite{Florencio85,Lee83, Lee82,Lee84}. To study the diffusive properties of the operator $A(t)$, we define the response function
\begin{equation}
\label{cor}
R(t)=\frac{ C_A(t)}{ C_A(0)},
\end{equation}
where
\begin{equation}
C_A(t)= \left\langle A(t) A(0) \right\rangle. 
\label{corr}
\end{equation}
We can analyze the diffusive behavior of the operator $A(t)$ by defining
\begin{equation}
 x(t)=\int_0^t A(s)ds,
\end{equation}
which allows us to define a diffusive behavior for $x(t)$. Now consider Eq.~\eqref{eq.sqr_disp}, replacing $D$ with $D(t)$, where
\begin{equation}
\label{DD}
D(t)= \int_0^{t}C_A(t')dt'.
\end{equation}

Note that $D(t)$ may not converge to a constant as in the Langevin case. The dispersion in the asymptotic regime, $t\to\infty$, is given by
\begin{equation}
\label{X2anomalous} 
\sigma_x^2 = \left\langle (x- \left\langle x\right\rangle)^2\right\rangle \sim D(t)t \sim t^\alpha,
\end{equation}
where $\alpha$ characterizes the type of diffusion. For $\alpha=1$, we have normal diffusion, distinguishing between subdiffusion $0<\alpha <1$ and superdiffusion $\alpha>1$. 

If we multiply Eq.~\eqref{GLE} by $A(0)$ and take the ensemble average, we obtain a self-consistent equation for $R(t)$
\begin{equation}
\label{self_consistent}
\frac{d R(t)}{d t}=-\int_0^t R(t-t')\Gamma(t') d t'.
\end{equation}

Once $R(t)$ is determined, the full system dynamics can be described. This approach represents a significant advancement from Eq.~\eqref{GLE} to \eqref{self_consistent}. With Eq.~\eqref{GLE}, many trajectories must be simulated to obtain meaningful results. However, with Eq.~\eqref{self_consistent}, we only need to solve an equation, which may have analytical solutions, as we will explore below. 

First, consider the nature of the noise, which can be associated with a set of harmonic oscillators of the form~\cite{Morgado02,Hanggi90}
\begin{equation}
\label{Noise0}
 F(t)=\left( 2 \langle A^2 \rangle_{eq}\right)^{1/2}\int_0^\infty \sqrt{ \rho(\omega)}\cos[\omega t+\phi(\omega)]d \omega,
\end{equation}
where the random phases, $0 \leq \phi(\omega) \leq 2 \pi$, provide the stochastic character. Inserting $F(t)$ into Eq.~\eqref{fdt}, we obtain
\begin{equation}
\label{memory}
\Gamma (t)=\int \rho (\omega )\cos (\omega t)d \omega ,
\end{equation}
where $\rho(\omega)$ is the noise density of states. The memory is even for any noise distribution~\cite{Vainstein05}. Note that for a classical system $\langle A^2 \rangle_{eq} \propto T$, and the noise, Eq.~\eqref{Noise0} is due to thermal fluctuations. Additionally,
\begin{equation}
\frac{d R(t)}{d t}\biggr|_{t=0}=0,
\end{equation}
as expected from Eq.~\eqref{self_consistent}. For noise density, consider
\begin{equation}
\label{noise_dos}
\rho (\omega)=
\begin{cases}
\frac{2\gamma }{ \pi } \left(\frac{\omega }{\omega_1} \right)^\nu, & \text{ if } \omega<\omega_1\\
\frac{2\gamma }{ \pi } , & \text{ if } \omega_1 <\omega<\omega_2\\
0, &\text{ if } \omega > \omega_2.
\end{cases}
\end{equation}

The constant in the density is chosen such that for normal diffusion, $\widetilde{\Gamma}(z=0)=\gamma$, as we will see below. This noise spectrum includes most of the noise discussed in the literature \cite{Morgado02,Costa03,Vainstein05,Vainstein06a,Ferreira12}. If $\omega_1=\omega_2$, it becomes a Debye cutoff frequency. We now consider three kinds of noise:\\

$(i)$ For $\nu=0$, we have~\cite{Morgado02}
\begin{equation}
\label{Me1}
\Gamma(t)=\frac{2\gamma \omega_2 }{ \pi }\left[\frac{\sin(\omega_2t) }{ \omega_2t } \right],
\end{equation}
which yields normal diffusion~\cite{Morgado02}. If $\omega_2 \rightarrow \infty$, we recover the white noise spectrum and $\Gamma(t) \rightarrow 2\gamma \delta(t)$. The Laplace transform of Eq.~\eqref{Me1} gives
\begin{equation}
 \widetilde{\Gamma}(z)=(2\gamma/\pi)\arctan(\omega_2/z),
\end{equation}
where the tilde denotes the Laplace transform.

A new timescale $t_2=\omega_2^{-1}$ arises naturally. For $t\gg t_2$, equivalent to $\widetilde{\Gamma}(z\rightarrow 0^\pm )= \pm\gamma$, $R(t)$ does not deviate from $R(t)=\exp(-\gamma |t|)$, indicating a memoryless Langevin equation. However, the short time behavior, $0 < t < t_2$, is quite different.\\ 

$(ii)$ For $\nu=1$ and $\omega_1=\omega_2$, we get~\cite{Ferreira12}
\begin{equation}
\label{Me2}
\Gamma(t)=\left(\frac{2\gamma \omega_1}{ \pi }\right)\left[\frac{\sin(\omega_1t)}{\omega_1t}+\frac{1-\cos(\omega_1t)}{(\omega_1t)^2} \right],
\end{equation}
with the Laplace transform
\begin{equation}
\label{LT2}
\widetilde{\Gamma}(z)=\left(\frac{\gamma z}{ \pi \omega_1 }\right)\ln\left[1+\left(\frac{\omega_1}{z}\right)^2 \right].
\end{equation}
For $\lim_{z \rightarrow 0} \widetilde{\Gamma}(z)\propto -z\ln{z}$, resulting in weak ballistic behavior~\cite{Ferreira12}. \\ 

$(iii)$ For $\nu \rightarrow \infty$, we get
\begin{equation}
\label{Me3}
\Gamma(t)=\left(\frac{2\gamma }{ \pi t }\right)\left[\sin(\omega_2t)-\sin(\omega_1t) \right],
\end{equation}
which yields $\lim_{z \rightarrow 0} \widetilde{\Gamma}(z) \propto z$, indicating ballistic diffusion~\cite{Morgado02,Costa03}. These memory functions are not artificial constructs; for example, Eq.~\eqref{Me3} was successfully used to describe spin wave diffusion in a random-exchange Heisenberg chain~\cite{Vainstein05,Vainstein05a}.

The motivation for considering such cases will become evident below. 

\begin{figure}[!h]
\centering
\includegraphics[width=\linewidth]{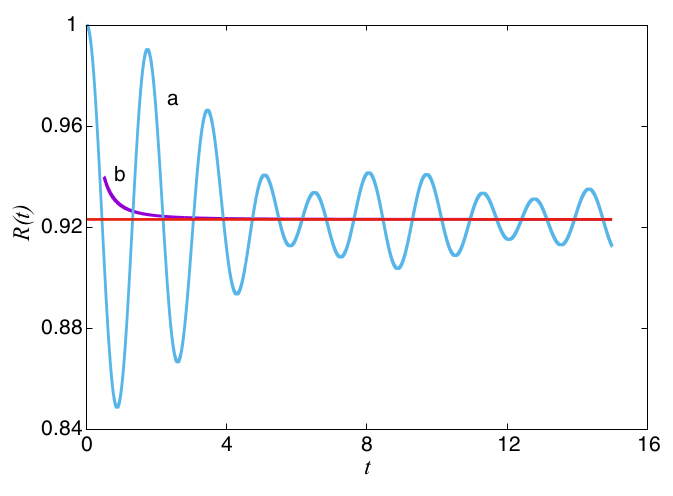}
  \caption{ Correlation function $R(t)$ as a function of $t$ in arbitrary units, {using the memory function~(\ref{Me3}) with $\omega_1 = 1$, $\omega_2 = 2$, and $\gamma = \frac{2}{\pi}$.  Curve (a) corresponds to the numerical integration of Eq.~\eqref{self_consistent}, and curve (b) to the scaling~(\ref{z1})-(\ref{ft}). The horizontal line represents the limit as 
  $t \rightarrow \infty$.  From Ferreira ref.~\cite{Ferreira22}, with permission from PRE.}}
\label{fig:corre}
\end{figure}

\begin{figure}[!h]
\centering
\includegraphics[width=\linewidth]{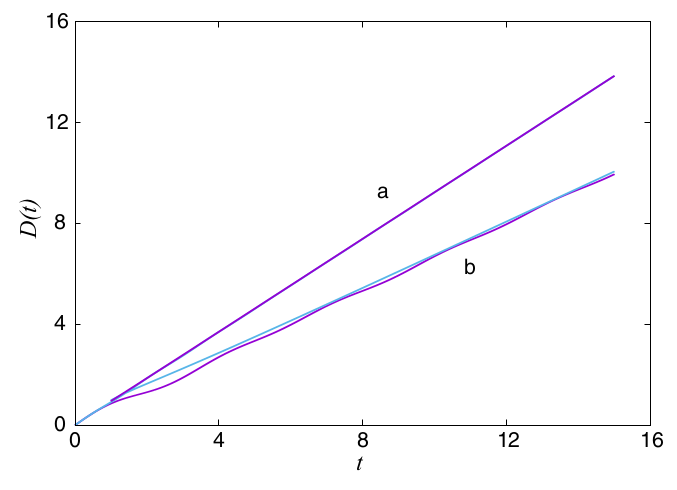}
 \caption{  The diffusion coefficient $D(t)$ as a function of time $t$. The curve with {small} oscillations {represents} the result obtained by the numerical
integration, {while} the less sinuous line corresponds to the
scaling. In both cases, $\gamma=\frac{2}{\pi}$.  {For} curve (a), $\omega_1=3$ {and} $\omega_2=4$.
{For} curve (b), $\omega_1=1$ {and} $\omega_2=2$. {Note that in (a)}, two curves overlap. From Ferreira ref.~\cite{Ferreira22}.}
\label{fig:diff}
\end{figure}

\subsubsection{Exact solutions for anomalous diffusion}

The major goal in any field of physics is to obtain exact results. In this section we display the exact exponents for anomalous diffusion.

Starting from Eq.~\eqref{DD}, we have
\begin{equation}
\label{Kubo2}
\lim_{t \rightarrow \infty}D(t)=\lim_{z \rightarrow 0} \int_0^{t}C_A(t')\exp(-zt')dt'=C_A(0)\lim_{z \rightarrow 0} \tilde{R}(z).
\end{equation}
In the last part, we use the final-value theorem~\cite{Morgado02}. Considering that $z \propto 1/t$, we can determine the asymptotic behavior of $D(t)$.

Applying the Laplace transform to Eq.~\eqref{self_consistent}, we obtain
\begin{equation}
\label{laplace_R}
\widetilde{R}(z)=\frac{1}{z+\widetilde{\Gamma }(z)}.
\end{equation}

Further analysis of this equation provides insights into the system's asymptotic behavior. Knowing $\widetilde{\Gamma}(z)$ allows us to predict the resulting dynamics. For instance, if $\lim_{z \rightarrow 0} \widetilde{\Gamma}(z) \propto z^\mu$, then the anomalous diffusion $D(t)$, Eq.~\eqref{Kubo2}, becomes
\begin{equation}
\label{diff_beta}
\lim_{t \rightarrow \infty}D(t) \propto t^\beta,
\end{equation}
where the diffusion exponent $\alpha$ is given by~\cite{Morgado02}
\begin{equation}
\label{alphaeq}
\alpha=\beta+1,
\end{equation}
with
\begin{equation}
\label{expbeta}
\beta=
\begin{cases}
\mu, &\text{ if~~ } -1<\mu <1\\
1, &\text{ if~~ } \mu \geq 1.\\
\end{cases}
\end{equation}

Thus, the exponent $\beta$ is constrained to a specific range, implying that the diffusion exponent $\alpha$ is also limited.

Most studies~\cite{Morgado02,Metzler00,Metzler04,Morgado04} have analyzed only the asymptotic behavior of diffusion as a power law described by Eq.~\eqref{X2anomalous}. Ferreira \textit{et al.}~\cite{Ferreira12,Ferreira22} proved that $z \propto 1/t$ as $t \rightarrow \infty$ is exact. More precisely, they define the scaling
\begin{equation}
\label{z1}
 z=\frac{\lambda(t)}{t},
\end{equation}
and proved that
\begin{equation}
\label{lambAna}
 \lambda(t)=1-\lim_{t \rightarrow \infty}t \frac{d}{dt}\ln{[f(t)]},
\end{equation}
with
\begin{equation}
\label{ft}
 f(t)=1-t\tilde{\Gamma}\left(\frac{\lambda(t)}{t}\right).
\end{equation}
These relations apply not only for large values of $t$ but also for intermediate values~\cite{Ferreira22}.

Figure~\ref{fig:corre} shows the correlation function obtained using the memory Eq.~\eqref{Me3}. The oscillating curves are obtained by numerical integration of Eq.~\eqref{self_consistent}, followed by Eq.~\eqref{DD}. The non-oscillating curves are obtained using Eq.~\eqref{lambAna}. Note the violation of the mixing condition, i.e. $R(t \rightarrow \infty) \neq 0$ (see Subsection~\ref{subsec:kte}).

Figure~\ref{fig:diff} shows the evolution of $D(t)$ using the correlation function obtained in Figure~\ref{fig:corre}. It is evident that $D(t)$ increases over time without converging to a fixed limit, indicating superdiffusive behavior.

They also considered a generalization of Eq.~\eqref{X2anomalous} as
\begin{equation}
\label{X2anomalous2}
\lim_{t \rightarrow \infty} \left\langle r^2(t)\right\rangle \sim t^\alpha \left[\ln (t) \right]^{\pm n}
 \sim t^{\alpha^{\pm}}.
\end{equation}

This definition for the $\alpha^{\pm}$ exponents was inspired by the critical exponents of phase transitions. For example, in magnetic systems with a null external field $H=0$, for temperatures $T \approx T_c$, the specific heat behaves as $C_{H=0} \propto|T-T_c|^{-\alpha}$, where $\alpha$ is the critical exponent. As is well known in the Ising model, in two dimensions $ C_{H=0} \propto \ln{|T-T_c|}$, leading to $\alpha = 0^{+}$.

This generalization illustrates the broad scope of diffusive behavior. As an example, for $t \rightarrow \infty$, an anomalous behavior of the form $\langle x^2(t) \rangle \propto \frac{t}{\ln(t)}$ was proposed by Srokowski~\cite{Srokowski00,Srokowski13,Oliveira01}. In our notation, this corresponds to $\alpha = 1^{-}$, representing a weak form of subdiffusion. In this context, the diffusive exponents can be categorized as follows
\begin{equation}
\label{diff_exponent}
\alpha= 
\begin{cases}
2, &\text{ if~~ } \nu>1\\
2^{-}, &\text{ if~~ } \nu=1\\
1+\nu, &\text{ if~~ } -1<\nu<1.
\end{cases}
\end{equation}

We thus identify two types of ballistic motion, with diffusion exponents $\alpha=2^{-}$ and $\alpha=2$. The weak ballistic motion $\alpha=2^{-}$ is ergodic, while the ballistic case $\alpha=2$ is not~\cite{Lapas08}. Although this difference is not easily observable in numerical simulations, a proper analytical analysis reveals it systematically~\cite{Lapas07,Lapas08,Ferreira12}. Note that all diffusion discussed here occurs in the absence of external fields, which would alter the results. 

This formalism can also be applied to study heat transfer between nanoparticles separated by a distance on the order of a few nanometers. Thermal fluctuations can excite surface waves (electromagnetic eigenmodes of the surface) inside a body. The intensity of surface waves is many orders of magnitude larger in the near-field than in the far-field, being quasi-monochromatic near the surface~\cite{Joulain05}. For near-field interactions, thermal conductance can be calculated assuming that both nanoparticles behave as effective dipoles at different temperatures~\cite{Domingues05}. Since these dipoles undergo thermal fluctuations, the FDT provides the energy that dissipates as heat in each nanoparticle. In a more general framework~\cite{Perez-Madrid08}, thermal conductance can be computed by assuming that the particles have charge distributions characterized by fluctuating multipole moments in equilibrium with heat baths at different temperatures. This quantity follows from the FDT for the fluctuations of the multipolar moments.

\subsubsection{Fluctuation–dissipation relation in anomalous transport systems}

An important contribution to the understanding of FDRs in anomalous transport systems was provided by Villamaina \textit{et al.}~\cite{villamaina2008}, who studied a granular single-file diffusion model consisting of a hard rods gas interacting via elastic or inelastic collisions on a ring, within the framework of Eq.~\eqref{L}. To evaluate the response function:
\begin{equation}\label{eq:rstar}
  R^*(t) = \frac{\langle \delta v(t) \rangle }{\delta v} = \frac{ \overline{ \left[ v'(t) - v(t) \right]} }{\left[ v'(t_0) - v(t_0)\right]},
\end{equation}
the authors implemented a cloning protocol that allows direct numerical measurement of the linear response to an impulsive perturbation. Here $v(t)$ and $v'(t)$ denote the tracer velocities in the unperturbed and perturbed systems, respectively. Once the system reaches thermalization, it is duplicated: one copy evolves unperturbed, while the other receives a small velocity increment $\delta v$, with $\delta v \ll \sqrt{T_g}$, applied to a tagged particle at time $t_0$; the kinetic temperature $T_g$, referred to as the granular temperature, is measured once the system reaches a statistically stationary regime. Both replicas are then evolved under the same realization of noise, and the response is computed as in Eq.~\eqref{eq:rstar} indicate the time dependent average in the dynamical ensemble generated by the external perturbation. We remark that the notation $R^*(t)$, used here to follow Villamaina \textit{et al.}, differs from the $R(t)$ adopted in our formalism (see Eq.~\eqref{cor}). While both capture the essence of FDRs, their methodological context differs.

An extended Einstein relation\footnote{While the original Einstein relation refers to the proportionality between diffusion and mobility at equilibrium, such as in Subsection~\ref{subsec:fluctuations_out_of_equilibrium}, modern usage -- particularly in fluctuation-response studies -- often extends the term to describe the dynamical identity between response functions and correlation functions under specific conditions~\cite{villamaina2008,Chaudhuri12}.} is specifically used to denote the time-dependent identity between the linear response function and the normalized velocity autocorrelation function, which can be expressed as $R^*(t) = R(t)$, where $R(t)$ corresponds to Eq.~\ref{cor} with $A = v$, and with $C_v(0)$ identified as the granular temperature.

Despite exhibiting sub-diffusive dynamics characterized by $\alpha = 1/2$, for which Eq.~\eqref{DD} is more appropriate, Villamaina \textit{et al.} showed that the Einstein relation holds in elastic systems or in regimes where the stationary phase-space distribution factorizes and the velocity statistics remain approximately Gaussian. However, in dense or strongly inelastic regimes, the emergence of inter-particle correlations and the breakdown of statistical independence lead to a violation of the Einstein relation. In such cases, the stationary phase-space distribution no longer factorizes, and only a more general form of the FDR remains valid. The Einstein relation is recovered under weak interaction conditions, such as low density or rapid thermalization.

Recent extensions of this formalism to active matter systems, where particles are subject to persistent self-propulsion and inherently driven out of equilibrium, have been developed by Caprini \textit{et al.}~\cite{Caprini18}. In these systems, the generalized FDR remains applicable by introducing effective response and correlation functions that explicitly account for self-propulsion forces. Their results reinforce the idea that the validity of Einstein-like relations is not strictly tied to equilibrium or normal diffusion, but rather hinges on the statistical structure of the phase space and the presence (or absence) of detailed balance.

In other words, these results emphasize that it is not anomalous diffusion itself that invalidates the FDRs, but rather the development of structural correlations and deviations from equilibrium-like statistics. Nonetheless, even when the Einstein relation fails, the system still operates within the linear response regime. These observations indicate that anomalous diffusion \textit{per se} does not necessarily imply a breakdown of the FDRs; rather, it is the loss of statistical factorization and the onset of collective correlations that undermine the validity of equilibrium-like response relations, as we shall see below from Kubo’s perspective (for further discussion, see Section~\ref{sec:anomalousdiff}).

\subsection{Limits of the Kubo formalism} \label{sec:violation}

The Kubo formalism is a powerful and exact method for determining various material properties and transport coefficients, such as conductivity, from appropriately chosen autocorrelation functions. It has been widely applied across numerous fields. However, it is important to explore the limitations of this framework by addressing a few general questions.

\subsubsection{Entropy and the second law of thermodynamics}

The first question concerns the validity of the second law of thermodynamics for stochastic processes. While we generally expect the second law to hold universally, can we prove it within the stochastic formalism presented here?

This question has been raised and discussed by several authors (see, e.g.,~\cite{Lapas07,Seifert12,Dybiec17}). In Ref.~\cite{Lapas07}, the total variation of entropy $\Delta S$ was calculated for a diffusion process and linked to the behavior of the memory kernel $\Gamma(t)$. Consider the scenario where:
\begin{equation}
 \tilde{\Gamma}(z\rightarrow 0) \rightarrow bz^\nu,
\end{equation}
with the effective values of $\nu$ ranging from $-1 \leq \nu \leq 1$. For $-1 \leq \nu < 1$, the entropy variation $\Delta S$ during the process is automatically positive. For $\nu = 1$, the condition is that
\begin{equation}
 b=\int_0^\infty \frac{\rho(\omega)}{\omega^2}d\omega >0,
 \label{eq:b}
\end{equation}
where $\rho(\omega)$ is the spectral density in Eq.~\eqref{memory}. Given that $\rho(\omega)>0$, the two equations above clearly indicate that the positivity of $\Delta S$ over a long time is guaranteed, thus upholding the second law of thermodynamics. Notably, a canonical form of the heat reservoir (a system of linearly coupled oscillators) responsible for the expression Eq.~\eqref{eq:b} is sufficient to ensure the second law's validity. 

\subsubsection{Khinchin's theorem, ergodicity, and violation of the fluctuation-dissipation theorem}\label{subsec:kte}

The concept of ergodicity, which has significant implications in statistical physics, was first introduced as a hypothesis by Boltzmann~\cite{Boltzmann74}. The ergodic condition, later formalized by Birkhoff and von Neumann for dynamical systems generating a flow in phase space, asserts that the time average and the ensemble average taken over all possible states with the same energy are equal, except for a negligible set of states. A key theorem related to the ergodic hypothesis is Khinchin's theorem~\cite{Burov10,Khinchin49,Lee07a}.

Khinchin's theorem states that, for a classical system, the ergodicity of a dynamic variable $A(t)$ in thermal equilibrium is maintained if the autocorrelation function Eq.~\eqref{corr} satisfies
\begin{equation}
\lim_{t\rightarrow \infty}\left\langle A(t) A(0) \right\rangle=\left\langle A\right\rangle \left\langle A\right\rangle.
\end{equation}

It is important to note that the theorem also establishes a relationship between ergodicity and the mixing condition. The mixing condition can be expressed as
\begin{equation}
\label{mixing}
 \lim_{t \rightarrow \infty}R(t)=0,
\end{equation}
which implies that, after a long time, the system forgets its initial conditions. According to Khinchin's theorem, if the mixing condition holds, then the Boltzmann ergodic hypothesis (BEH) also holds.

In recent decades, researchers have explored the validity of Khinchin's theorem and ergodicity in systems described by the Mori formalism~\cite{Costa03,Lapas07,Lapas08,Lee07a,Lee01,Bao25}. For example, consider the limit
\begin{equation}
\label{Mix2}
\lim_{t \rightarrow \infty}R(t)=\lim_{z \rightarrow 0}z \tilde{R}(z).
\end{equation}
For systems described by the GLE, $\tilde{R}(z)$ is given by Eq.~\eqref{laplace_R}, and if we assume that as $z\rightarrow 0$ the leading term in $\tilde{\Gamma}(z)$ behaves as $\tilde{\Gamma}(z) \rightarrow bz^{\alpha-1}$, then we can write
\begin{equation}
\label{Mix3}
\Lambda=\lim_{t \rightarrow \infty}R(t)=\lim_{z \rightarrow 0}\frac{1}{1+bz^{\alpha-2}}.
\end{equation}
It becomes evident that, for $\alpha<2$, the mixing condition holds, and thus the BEH holds according to Khinchin's theorem. For ballistic motion $\alpha = 2$, $\Lambda=(1+b)^{-1} \neq 0$, and the mixing condition does not hold. Consequently, the BEH fails. The factor $\Lambda$ is therefore referred to as the non-ergodic factor~\cite{Costa03}.

We can express the solution of Eq.~\eqref{GLE}, analogous to the Langevin equation, Eq.~\eqref{L}, as
\begin{equation}
\label{At}
A(t)=A(0)+\int_0^t F(s)R(s)ds.
\end{equation}

Squaring and taking the time average yields
\begin{equation}
\label{At2}
\overline{ A^2(t)}=\langle A^2 \rangle_{eq}+R(t)^2\left[\langle A^2(0)\rangle-\langle A^2\rangle_{eq} \right].
\end{equation} 

If the system is initially at equilibrium $\langle A^2(0)\rangle=\langle A^2\rangle_{eq}$, it will remain in equilibrium indefinitely, i.e., no fluctuation will drive it away from equilibrium. However, if the system is initially out of equilibrium $\langle A^2(0)\rangle \neq \langle A^2\rangle_{eq}$, it will approach equilibrium as $R(t \rightarrow \infty)=0$, indicating that the mixing condition Eq.~\eqref{mixing} is satisfied. Conversely, if the time average does not equal the ensemble average, the BEH fails~\cite{Costa03}. Thus, we arrive at an important conclusion:
\begin{itemize}
 \item If $\Lambda=0$, the mixing condition holds, and so does the BEH.
 \item If $\Lambda \neq 0$, the mixing condition does not hold, and the BEH also fails.
\end{itemize}

In all of the above cases, Khinchin's theorem remains valid.

Using equations similar to Eqs.~\eqref{At} and~\eqref{At2}, we can also derive higher-order moments and define the non-Gaussian factor
\begin{equation}
\eta(t)= \left[ \frac{\langle A^2(0)\rangle}{\langle A^2(t)\rangle} \right]^2 \eta(0)R^4(t),
\end{equation}
and the skewness parameter~\cite{Lapas07,Lapas08}:
\begin{equation}
\phi(t)= \left[ \frac{ \sigma_A(0)}{\sigma_A(t)} \right]^3 \phi(0)R^3(t).
\end{equation}

The non-Gaussian factor measures the deviation of the distribution of a stochastic variable $A(t)$ from Gaussian statistics. Note that if $\eta(0)=0$, the distribution will always be Gaussian. The skewness parameter $\phi(t)$ has a similar property. For instance, if the initial $\phi(0)$ is zero, the system's dynamics will preserve the symmetry of the distribution.

If the initial distribution is non-Gaussian ($\eta(0) \neq 0$, $\phi(0) \neq 0$), both indicators evolve towards a more symmetrical final distribution as $t \rightarrow \infty$. This is demonstrated by the ratio $\eta(t)/\eta(0)$, which is less than 1 as long as $R^2(t)< 1$. Moreover, since $\lvert R(t) \rvert <1$, the process of ``Gaussianization'' occurs much faster than the decay of $R(t)$. For example, if $R(t)/R(0) \approx 10^{-1}$, then $\eta(t)/\eta(0) \approx 10^{-4}$. Additionally, if the mixing condition is fulfilled, the long-time distribution becomes Gaussian. 

As discussed, the relationship between fluctuations and dissipation plays a crucial role in response theory based on statistical mechanics. The FDT formalism, derived for near-equilibrium systems where detailed balance holds, allows us to obtain important measured quantities such as susceptibility, the light scattering cross section, neutron scattering intensity, diffusion, surface roughness in growth, etc., by studying spontaneous correlations of fluctuating dynamic variables.

On the other hand, the breakdown of the equilibrium FDT is a well-documented phenomenon in the literature. For Kardar-Parisi-Zhang (KPZ) dynamics in $d+1$ dimensions, the FDT holds for $d=1$, but it fails~\cite{Kardar86,Rodriguez19} for $d >1$, see Section~\ref{Sec_FDTKPZ}. For phase transition, it works only for the upper critical dimension~(see Section. \ref{sec:phase}). For these two cases, the corrections involve moving from an Euclidean to a fractal dimension. Furthermore, violations of the FDT have also been observed in structural glass experiments through extensive computer simulations~\cite{Barrat98,Bellon05,Bellon02,Crisanti03,Grigera99,Ricci-Tersenghi00}, in proteins~\cite{Hayashi07}, in mesoscopic radiative heat transfer~\cite{Perez-Madrid09,Averin10}, as well as in ballistic diffusion~\cite{Costa03,Lapas07,Lapas08}.

One step toward understanding the scenarios in which the FDT breaks down is to analyze a hierarchy of the fundamental theorems of statistical physics: the mixing condition is stronger than the BEH, and ergodicity is a necessary condition for the FDT, as discussed above. The hierarchical connection among mixing, ergodicity, and FDT has been investigated in a series of works~\cite{Costa03,Lapas07,Lapas08,Lee07a,Lapas15a} using Lee's recurrence relations, thereby establishing the way in which the FDT relation may be contravened.

\subsubsection{Anomalous relaxation}

Kubo's theory of time-dependent correlations is based on Onsager's regression hypothesis, which assumes that the relaxation of a perturbed macroscopic system follows the same law that governs the dynamics of fluctuations in equilibrium systems~\cite{vanKampen92,Kubo1957,Zwanzig01,Mori65}. Near equilibrium, the FDT describes a relationship linking relaxation to correlations between spontaneous fluctuations in the system. In the linear regime, the deviation of a dynamic variable $A(t)$ from its initial value caused by a time-dependent perturbation $h(t)$ is given by
\begin{equation}
 \Delta A(t,\epsilon h(t)) = \epsilon \Delta A(t,h(t)),
\end{equation}
with the general solution
\begin{equation}
\Delta A(t)= \int^{\infty}_{-\infty}dt'R(t,t')h(t').
\end{equation}
In addition,
\begin{equation}
   R(t)=-\text{const}\frac{d}{dt}\left\langle \delta A(0)\delta A(t)\right\rangle,\qquad t>0.
\end{equation}

As an example, let $A(t)$ be proportional to the concentration of species $c_A(t)$ undergoing a first-order chemical reaction $A\leftrightarrow B$ with detailed balance secured. As a result,
\begin{equation}
\frac{\left\langle \delta A(0)\delta A(t)\right\rangle}{\left\langle \delta A_{eq}^2\right\rangle}\propto \exp(-t/\tau_{relax}),
\end{equation} 
with $\tau_{relax}$ representing the relaxation time related to the inverse of the kinetic rate $\tau_{relax}=(k_{AB}+k_{BA})^{-1}$. The decay of the concentration of species over time to its stationary (equilibrium) value can also be described by the survival probability, i.e., the probability that the amount (concentration) of species has not reacted until time $t$.

The first reported relaxation law was Newton's law of cooling, which, for small temperature differences between the body and its environment, predicts that the rate of cooling of a warm body at any moment is proportional to the temperature difference between the body and its surroundings. Accordingly, the temperature of the cooling body with respect to the environment decreases exponentially over time. However, such an exponential relaxation law is often a rough approximation of the various physical mechanisms responsible for the relaxation process. 

An extensive literature on relaxation kinetics in complex systems refers to the plethora of observed non-exponential decays of correlation functions, such as those seen in growth phenomena~\cite{Colaiori01}, supercooled colloidal systems~\cite{Rubi04}, hydrated proteins~\cite{Peyrard01}, glasses and granular materials~\cite{Santamaria-Holek04,Vainstein03a}, disordered vortex lattices in superconductors~\cite{Bouchaud91}, plasma~\cite{Ferreira91}, and liquid crystal polymers~\cite{Santos00,Benmouna01}. Such systems exhibit physical properties similar to those observed in particle motion described by continuous time random walks and anomalous diffusion~\cite{Vainstein05,Costa06,Lapas15}. The quest to obtain response functions that can explain such relaxation processes has been an ongoing subject of study for over a century.

Rudolph Kohlrausch used stretched exponentials $R(t) \approx \exp{[-(t/\tau)^{\beta}]}$ with $0 < \beta < 1$ to describe charge relaxation in a Leyden jar~\cite{Kohlrausch54}. Later, his son, Friedrich Kohlrausch~\cite{Kohlrausch63}, observed two distinct universalities: the stretched exponential with $0 < \beta < 2$ and the power law dependence. The former behavior is now known as the Kohlrausch-Williams-Watts stretched exponential. The main methods used to describe these empirical relaxation patterns are similar to those explored in studies of anomalous diffusion. 

For instance, for an even response function $R(-t)=R(t)$, the time derivative $\frac{dR(t)}{dt}$ must be zero at $t=0$, which contrasts with the result $R(t)=\exp(-\gamma \vert t\vert)$ from the memoryless Langevin equation~\cite{Vainstein06}. However, it can be shown that the exponential is a reasonable approximation in some cases. Vainstein {\it et al.}~\cite{Vainstein06a} discussed various forms of correlation functions that can be obtained from Eq.~\eqref{self_consistent} once $\Gamma(t)$ is known, such as the Mittag-Leffler function~\cite{Mittag-Leffler05}, which behaves as a stretched exponential at short times and adopts an inverse power law form in long-time regimes. Even for the simplest case of normal diffusion, $\nu=0$ (Eq.~\eqref{Me1}), $R(t)$ is not an exponential since its derivative at the origin is zero; however, for broad-band noise $\omega_s \gg \gamma $, i.e., in the limit of white noise, it approaches the exponential $R(t)= \exp(-\gamma t)$ for times greater than $\tau_s= \omega_s^{-1}$. 

Recent advances in stochastic modeling of anomalous kinetics observed in complex dielectric materials involving self-similar random processes have been broadly investigated in a series of papers by Weron and collaborators~\cite{WeronK,Novak05}.

\subsubsection{Fluctuation-dissipation theorem, anomalous diffusion, and generalized L\'{e}vy walk \label{sec:anomalousdiff}}

In this concluding part of the discussion, we return to the ideas presented in the introduction, particularly the CLT and the IRGT, Eq.~\eqref{Pnk}. This brought us to stable L\'{e}vy functions $L(\nu,x)$ defined in Eq.~\eqref{Levy}. In particular, for $\nu=1$, we observed a Lorentzian distribution, and for $\nu=2$, a Gaussian distribution curve.

Since L\'{e}vy's original mathematical works~\cite{Levy37}, L\'{e}vy statistics have become widely used. Non-Gaussian stochastic L\'{e}vy processes, i.e., processes with independent, stationary increments distributed according to a stable density with $\nu\neq 2$, serve as prototypes for describing unusual transports leading to anomalous diffusion~\cite{Metzler99,Metzler00,Metzler04}. Among this class, free L\'{e}vy flights (LFs) represent a special category of discontinuous Markovian processes, for which the MSD $\left\langle(\Delta x( t))^2\right\rangle$ diverges due to the heavy-tail distribution of the independent increments $\Delta x$. Unlike L\'evy walks (LWs), which are non-Markovian processes characterized by spatiotemporal coupling, LF corresponds to Markov processes emerging from a Langevin equation with delta-correlated, white L\'evy noise. It is important to note that when an overdamped particle is subjected to white L\'evy noise, time symmetry is broken, and microscopic reversibility is violated, even when the noise is symmetric~\cite{Kusmierz2016,Ebeling2013}. 

In a series of papers~\cite{Dybiec12,Dybiec17,Kusmierz2016,Ebeling2013}, authors studied a one-dimensional Langevin model in which L\'evy fluctuations are incorporated either as a random external forcing or as a non-thermal bath. These scenarios clearly correspond to systems that are strictly non-equilibrium. Even though moments of force and energy transfer distributions in these systems do not exist, it can be shown that the FDT and linear response theory can still apply to properly defined conjugate variables. The generalized form of the FDT relies on Markovian dynamics and the definition of the stationary probability distribution $p_s(x,f(t))$, which represents the stationary state of the non-equilibrium system achieved in the presence of external forcing $f(t)$ and continuous dissipation. This approach can be effectively applied to models of non-Hamiltonian systems~\cite{Dybiec12,Marconi08,Sarracino19,Hanggi82}, with the generalized potential $V=-\ln p_s(x;f)$. By analogy with Onsager's theory, the variable conjugate to the perturbation $f(t)$ can be defined as
\begin{equation}
X=-\frac{\partial \ln p_s(x,f(t))}{\partial f}|_{f=0},
\end{equation}
and relates to the generalized susceptibility $\chi(t)=\frac{d}{dt}\left\langle X(t)X(0)\right\rangle_0$ in the linear response theory $\left\langle X(t)\right\rangle=\int_0^t\chi(t-s)f(s)ds$, where the index 0 refers to the unperturbed state. This generalization of the FDT shows that the conjugate variable represents the change in the probability distribution of the system under the perturbation. In equilibrium, this change is also related to the energy that the system absorbs from the perturbation. However, for non-equilibrium systems, particularly those driven by L\'evy noises, the lack of conserved quantities may prevent such an interpretation~\cite{Ebeling2010,Ebeling2011}. 

To avoid the divergence of the second moment in classical LFs, further modifications can be made by penalizing long LF jumps with extended waiting times, effectively introducing a finite velocity of motion~\cite{Metzler00,Metzler04,Zaburdaev15}. Jump lengths and waiting times for the next executed step may be linearly coupled, such that the resulting L\'evy walk proceeds in a given direction with constant speed until a velocity reversal after a waiting time, or the space-time coupling may have a more complex, power-law relationship. Also, the velocity of motion may change from one step to another. Interestingly, even within this construct to remedy the divergence of the second moment (or the MSD, LWs can still exhibit an infinite variance for certain parameters~\cite{Weron10,Thiel13,Dorea06,Dybiec12,Dybiec17,Zaburdaev16,Albers2018}. 

In a series of studies exploring LWs and their role in superdiffusive transport, it has been shown that the microscopic geometry of these random walks can significantly influence the resulting dispersal regimes. To illustrate this complex behavior, we consider the non-ergodicity of $ d $-dimensional generalized Lévy walks as described by Zaburdaev et al.~\cite{Zaburdaev16} and Albers and Radons~\cite{Albers2018}.

Zaburdaev et al.~\cite{Zaburdaev16} extended the concept of LWs to two dimensions, revealing that the asymptotic probability density functions (PDFs) of superdiffusive LWs reflect the underlying random walk structure. These findings provide a framework for understanding various transport regimes observed in nature, ranging from biological motility to the motion of cold atoms.

\begin{figure}[!h]
\centering
\includegraphics[width=\linewidth]{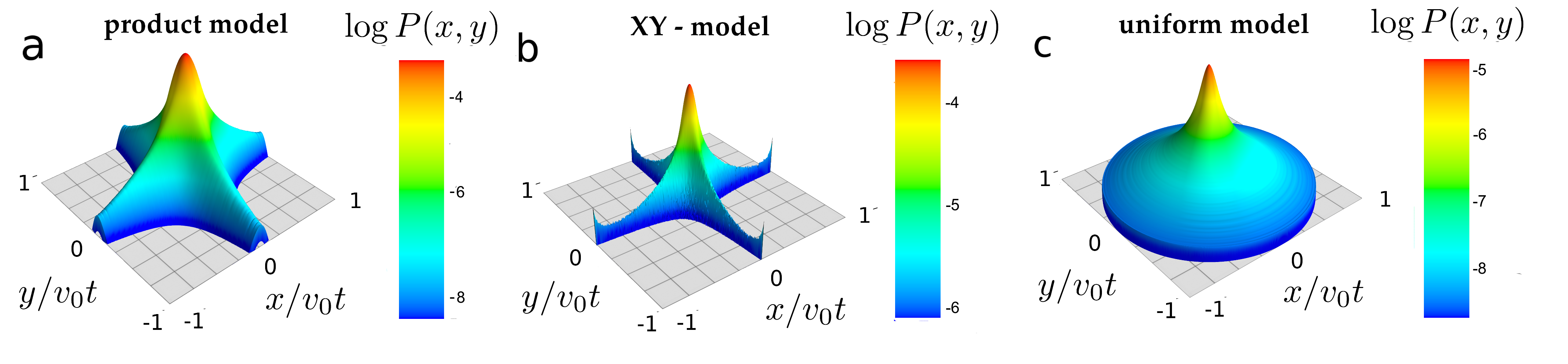}
 \caption{Probability density functions (PDFs) of the three models in the superdiffusive regime. The distributions are plotted on a log scale for the time $t/\tau_0 =10^4 $. The PDF for the product model (a) was obtained by multiplying PDFs of two identical one-dimensional LW processes. The PDFs for the XY (b) and uniform (c) models were obtained by sampling over $10^{14}$ realizations. The parameters are $\gamma=3/2$, $v_0=1$, and $\tau_0=1$.
 From ref.~\cite{Zaburdaev16}.}
\label{fig:prl2}
\end{figure}

The probability density function of a flight is given by~\cite{Zaburdaev16} 
\begin{equation}
  \psi(t)= \frac{\gamma }{t_0(1+t/t_0)^{1+\gamma}},
\end{equation}
with $\gamma>0$ and $t_0>0$. Depending on the value of $\gamma$, the LW exhibits different dispersal behaviors. It can lead to a dispersal $\alpha = 1$, typical for normal diffusion ($\gamma > 2$), and very long excursions leading to the fast dispersal with $1<\alpha\leq 2$ in the case of superdiffusion $0 < \gamma < 2$). Figure~\ref{fig:prl2} illustrates the probability density functions of three distinct two-dimensional LW models: the product model, the XY model, and the uniform model. These PDFs were obtained by sampling over $10^{14}$ realizations, clearly demonstrating the differences between each model in the superdiffusive regime. The product model is constructed by multiplying two independent one-dimensional LW processes, while the XY and uniform models exhibit distinct geometric characteristics. Notably, the XY model is constrained to motion along a single axis at a time, whereas the uniform model allows random motion in any direction. The superdiffusive nature of these processes, particularly within the sub-ballistic regime, is evident in the non-Gaussian tails of the PDFs, which deviate from the universal Gaussian form typical of standard random walks. As shown, these findings align with the theoretical framework established in Ref.~\cite{Zaburdaev16}, where the space-time coupling in LW gives rise to anomalously fast diffusion, as discussed in the previous section.

\begin{figure}[!h]
\centering
\includegraphics[width=\linewidth]{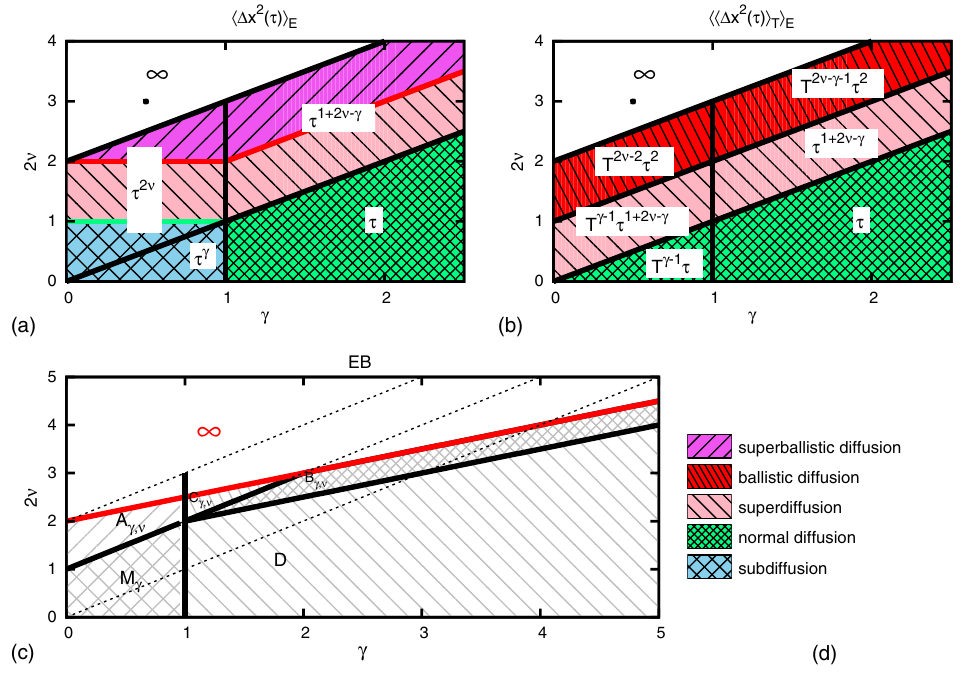}
 \caption{Phase diagram for the ensemble-averaged squared displacement (a), the ensemble average of the time-averaged squared displacement (b), and the ergodicity breaking parameter (c). Different ranges of validity of the analytical results are separated by thick black lines in the two-dimensional parameter space. Different kinds of diffusion are color-coded as indicated in the key. The dotted lines in (c) serve as a guide to the eye for a better comparison with the phase diagram in (b). The ergodicity breaking parameter EB is defined as the variance of the rescaled random variable EB$=\left\langle \hat{\xi}^2(\tau) \right\rangle_E-1$, where $\left\langle \hat{\xi}(\tau) \right\rangle = \left\langle \Delta x^2(\tau) \right\rangle_T/\left\langle \left\langle \Delta x^2(\tau) \right\rangle_T \right\rangle_E$ with subscripts referring to time (T) and ensemble (E) averages. Adapted from Albers and Radons~\cite{Albers2018}.}
 \label{fig:phasediagram}
\end{figure}

The absolute values $V_i$ of the velocities of the flights are constant during a single flight and depend on the flight duration in a deterministic way, $V_i \propto t^{\nu-1}$; $\nu>0$, where the flight directions are isotropically chosen at random. From the derived trajectories, one computes the ensemble and time-averaged squared displacement, as displayed in Figure \ref{fig:phasediagram}. Altogether, the space-time coupled LW is a generalized form of Brownian motion, combining two key features: the ability to generate anomalously fast diffusion and a finite velocity of a random walker. The motion exhibits nontrivial phenomena: the MSD of the generalized LW diverges for $2\nu\geq \gamma+2$. Within the parameter range $\gamma<1$ and $\gamma<2\nu<1$, the ensemble-averaged squared displacement exhibits subdiffusion, while the time-averaged squared displacement indicates superdiffusive motion. Notably, the motion can also reflect normal (linear in time) MSD growth for non-Gaussian diffusion.

In conclusion, the features of the modeled motion, including its weak ergodicity breaking, indicate potential difficulties and limitations when using such models for the statistical analysis of experiments based on single-particle tracking.

\subsubsection{L\'{e}vy walks and path chaos in vortical flows}

In exploring complex transport behaviors in fluid systems, the work by Hu et al.~\cite{Hu2021} provides insight into the dynamics of semi-rigid filaments in vortical flows, revealing a rich spectrum of behaviors ranging from LW to chaotic dispersal. Through both simulations and experimental observations, they demonstrate how fluctuations in the vortical field result in distinct transport regimes that depend heavily on filament length $L$, vortex size $W$, and effective flexibility $\eta$, transitioning from LW, of generally decreasing diffusion exponents and increasing Lyapunov exponents, to Brownian motion.

\begin{figure}[htbp]
  \centering
  \includegraphics[width=\linewidth]{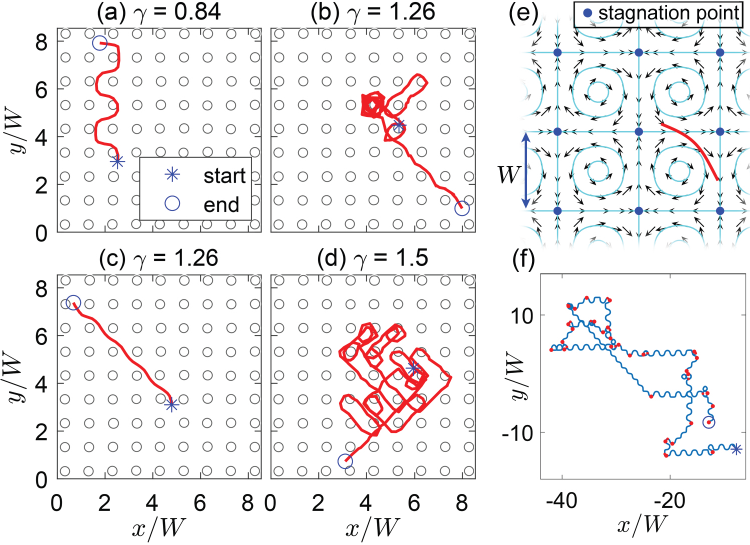}
  \caption{{Experimental center-of-mass (COM) trajectories of flexible ribbons in a 9-by-9 cellular flow, as shown on the left. Circles indicate the locations of the spinning rollers. (a) A strongly undulating step along the $+y$ direction with $\eta \approx 25$. (b) A meandering trajectory with a long undulating diagonal step with $\eta \approx 85$. (c) Diagonal step with $\eta \approx 85$. (d) Meandering trajectory with many turns with $\eta \approx 118$. (e) Snapshot from a numerical simulation with $\eta = 100$ and $ \gamma = 1$, showing a flexible filament moving in the cellular flow described by Eq.~\eqref{eq:Phi}. Black arrows represent the background cellular flow, and cyan closed curves are the streamlines. (f) A typical trajectory from a simulation with $\gamma = 1$ and $\eta = 0.5$. The red points are turning points separating two different steps. Figure from Hu et al~\cite{Hu2021}.}
}
  \label{fig:levy_walks}
\end{figure}

The experimental results, as depicted in Figure~\ref{fig:levy_walks}, highlight different filament trajectories in a 9-by-9 cellular vortical flow. Shorter filaments, such as those with $\eta \approx 25$, exhibit strongly undulating paths along the $+y$ direction, while longer filaments (e.g., $\eta \approx 85$ or $\eta \approx 118$) show more complex trajectories, including meandering and diagonal steps. These distinct motion types can be classified as LW, which follow power-law distributions of step-lengths, with frequent short steps interrupted by occasional long jumps~\cite{Zaburdaev15}.

The transition from LW to Brownian motion is driven by the filament's ability to sample the flow field at varying length scales. As the filament length increases relative to the vortex size (quantified by the ratio $\gamma = L/W$), the system evolves from exhibiting path chaos to more diffusive, Brownian-like behavior~\cite{Zaburdaev16}. This change can be understood through the flow's stream function 
\begin{equation} 
\Phi_\gamma = (\pi \gamma)^{-1} \sin(\pi \gamma x) \sin(\pi \gamma y), 
\label{eq:Phi}
\end{equation} 
which governs the background flow experienced by the filament and has stagnation points at $(n,m)\gamma^{-1}$ for $n$, $m$ integers.

The LW behavior exhibited by the filaments is closely related to the generalized LW studied by Zaburdaev et al.~\cite{Zaburdaev16}, where step lengths and waiting times are coupled in a power-law fashion. This coupling leads to superdiffusive transport, where the MSD grows faster than in normal diffusion processes. The intricate dynamics seen in Hu et al.'s vortical flow system provide a practical realization of these theoretical LW models. Even as modifications such as finite velocities or velocity reversals are introduced to prevent divergence in the MSD~\cite{Zaburdaev15, Zarfaty2019}, the resulting motion remains fundamentally non-Markovian and non-ergodic, offering a rich ground for further exploration.

As illustrated by the filament trajectories in Figure~\ref{fig:levy_walks}
(e-f), numerical simulations complement the experimental observations. For instance, a flexible filament with $\eta = 100$ and $\gamma = 1$ shows characteristic LW behavior, while a filament with $\eta = 0.5$ transitions to more diffusive motion. These findings align with the predictions of generalized L\'{e}vy walks, where space-time coupling leads to non-trivial transport dynamics, including superdiffusion and subdiffusion depending on system parameters~\cite{Zaburdaev15, Zaburdaev16, Zarfaty2019}.

\section{Fluctuation and growth} \label{sec:growth}

One of the most significant manifestations in nature is the phenomenon of growth. It is ubiquitous in biology, where it occurs in processes such as cell growth, population dynamics, and embryo development. It also plays a crucial role in chemistry, where reagents and products exchange their populations, and in physics, where the stability of growth or its absence is often in question. Fluctuations are inherent in these processes, and we will discuss two basic types:

\begin{enumerate}
\item Small fluctuations, such as those in population dynamics.
\item Large fluctuations, such as those in material growth.
\end{enumerate}

This selection covers a wide range of phenomena and represents very active areas of research.

\subsection{Small fluctuations and population dynamics}

As mentioned above, the concepts of diffusion and fluctuations have found applications in various branches of the natural sciences. Extending these concepts has immediate applications in the study of growth. Since growth phenomena are widely observed in various dynamic processes in nature, they have received significant attention in the study of dynamics over the past few years. Broadly speaking, we can categorize these studies into two types: those with small fluctuations, where they appear as a perturbation, and those with large fluctuations. As an example of the first type, let us consider the total population $N(t)$ in a given ecological niche. A simple equation to model this growth is given by
\begin{equation}
\label{Log}
\dfrac{d N(t)}{d t} = aN(t) - bN(t)^2,
\end{equation}
where $a$ and $b$ represent the growth rate and the competition rate, respectively. For $b = 0$, we have the Malthusian exponential growth. The above equation is known as the logistic growth of Verhulst~\cite{Banks94,Murray2002}, which imposes a limit for growth, $N(t \rightarrow \infty) = a/b$. In a more elaborate analysis, we are not only interested in the total number $N(t)$, but also in the local density $u(\vec{x},t)$, which propagates in time and space, as in the famous Fisher equation~\cite{Fisher37}
\begin{eqnarray}
\frac{\partial u(x,t)}{\partial t} = au(x',t) + D\frac{\partial ^2u(x,t)}{\partial x^2} - bu(x,t)^2,
\label{App3}
\end{eqnarray}
where $D$ is the diffusion coefficient. Here, $N(t) = \int u(\vec{x},t)d\vec{x}$. Note that the uniform density $u_0 = a/b$ is also a solution of the above equation. Fisher was the first to show that bacterial populations can propagate as a wave, now known as Fisher waves. This and subsequent studies suggested we can model the evolution of density using reaction-diffusion equations~\cite{Banks94,Murray2002,Rothe84}. Recently, a general reaction-diffusion equation has been proposed~\cite{Aranda20-1,Aranda20-2,Aranda21}
\begin{equation}
\label{Gen}
\dfrac{\partial u(\vec{x},t)}{\partial t} = \hat{M}u(x,t),
\end{equation}
where $\hat{M}$ is an operator acting on $u(\vec{x},t)$. In this form, Eq.~\eqref{Gen} is quite general and can describe a large number of different physical systems (or properties), such as the evolution of particle density~\cite{Landau65,Cross93}, energy, or even living organisms as in a colony of bacteria~\cite{Murray2002,Fisher37,Turing52,Kolmogorov37}.

For simple systems in which the operator $\hat{M}$ is linear, analytical solutions are readily obtained. However, most realistic systems are described by a nonlinear $\hat{M}$ (cf. Eqs.~\eqref{Log} and~\eqref{App3}), with the operator implicitly depending on $u$, $\hat{M} = \hat{M}(u)$; in these cases, analytical solutions are cumbersome and not easily obtained.

\subsubsection{Nonlocal growth}

Here we present an example of a one-dimensional system in the range $0 < x < L$, described by an equation that includes nonlocal growth and interaction terms involving long-range effects in the system. This equation takes the form~\cite{DaCunha11}
\begin{eqnarray}
\frac{\partial u(x,t)}{\partial t} = a\displaystyle{\int_{0}^{L}}g_{\alpha}(x-x')u(x',t)dx' - bu(x,t)\displaystyle{\int_{0}^{L}}g_{\beta}(x-x')u(x',t)dx',
\label{App2}
\end{eqnarray}
where $\alpha$ and $\beta$ are the growth and competition length parameters, respectively. A simple choice for the nonlocal kernel is
\begin{eqnarray}
\label{ERD_gkappa}
g_\kappa(x) = \left\{ \begin{array}{lcl}
\frac{1}{2\kappa}, & |x| < \kappa, & \\
0, &\text{otherwise}. & \\
\end{array}\right.
\end{eqnarray}

Despite all the complexity, there are some general characteristics for these systems. For instance, three questions need to be answered: First, does a uniform solution $\hat{M}u_0 = 0$ exist, i.e., the constant density $u_0 = a/b$? Note that Eq.~\eqref{App2} satisfies this condition. Second, is this uniform solution stable? And third, what happens when fluctuations disturb it? Landau proposed a simple and general answer in his studies of plasma stability~\cite{Landau65,Cross93}. For him, fluctuations were just small initial inputs added to the uniform solution $u_0$ in the form
\begin{eqnarray}
u(\vec{x},t) = u_0 + \epsilon \exp{\left[\gamma(k) t\right]}\cos{\left(\vec{k}\cdot\vec{x}\right)}, 
\label{pert}
\end{eqnarray}
where $\epsilon \ll u_0$ is a perturbation that, for large $t$, can either grow or decay, depending on the value of $\gamma(k)$. We then solve Eq.~\eqref{Gen} to first order in $\epsilon$ to obtain $\gamma(k)$, such that
\begin{align}\label{lambda}
\gamma(k) &< 0, \quad 
\text{
 \begin{tabular}{l}
  the system evolves to the homogeneous \\ 
  steady-state.
 \end{tabular}
} \\
\gamma(k) &> 0, \quad 
\text{
 \begin{tabular}{l}
  the system departs from the homogeneous state, \\
  leading to pattern formation.
 \end{tabular}
}
\end{align}

This approach has been widely used in nonlinear dynamics to study pattern formation~\cite{Oliveira96,Gonzalez99,Aranda20-1,Aranda20-2,DaCunha11,Koch94,Nicolis95,Newell97,Rabinovich20,Fuentes03,Fuentes04,DaCunha09,Colombo12,Barbosa17}. 

\begin{figure}[h]
\centering
\includegraphics[width=\linewidth]{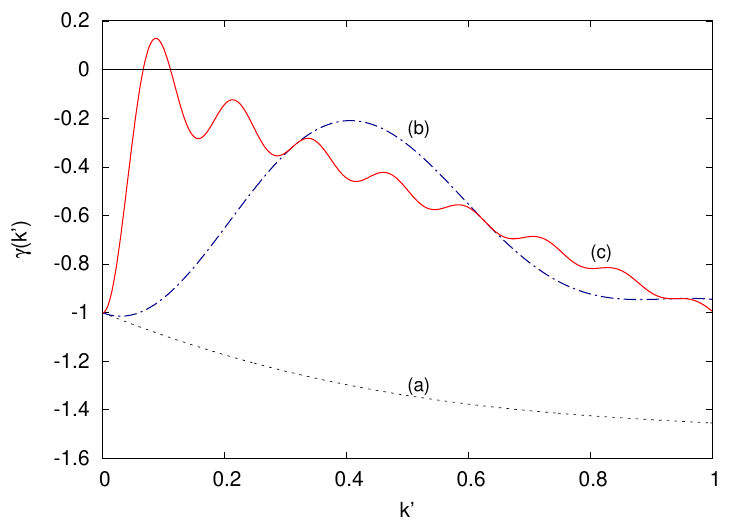}
\caption{The effective growth rate $\gamma(k')$ as a function of $k'$ for a small perturbation from the uniform density {is shown for different values of $\eta$:} (a) $\eta = 2$; (b) $\eta = 10$; (c) $\eta = 50$. For values of $\gamma(k') > 0$, the system departs from the homogeneous state, leading to pattern formation. From Fuentes~\textit{et al.}~\cite{Fuentes04}. }
\label{fig:gamma}
\end{figure}

As an example, Fuentes~\textit{et al.}~\cite{Fuentes04} used
\begin{eqnarray}
\frac{\partial u(x,t)}{\partial t} = au(x',t) + D\frac{\partial^2 u(x,t)}{\partial x^2} - bu(x,t)\int_{0}^{L}g_{\beta}(x-x')u(x',t)dx',
\label{AppF}
\end{eqnarray}
to obtain
\begin{eqnarray}
\label{lambd0}
\gamma(k') = -a\left[\frac{\sin\left( k'\eta \right)}{k'\eta} + k' \right], 
\end{eqnarray}
with $k' = k\sqrt{D/a}$ and $\eta = \beta\sqrt{a/D}$. This is illustrated in Figure~\ref{fig:gamma}, showing $\gamma(k')$ as a function of $k'$ for three different values of $\eta$: $50, 10$, and $2$. Here $\gamma$ is in units of $a$, and $k$ in units of $L^{-1}$. Fluctuations with values of $k'$ such that $\gamma(k') > 0$ will be amplified, while those with $\gamma(k') < 0$ will decay. We apply Landau's approach, Eq.~\eqref{pert}, to Eq.~\eqref{App2} to get
\begin{eqnarray}
\label{lambd}
\gamma(k) = a\left[\frac{\sin\left( k\alpha \right)}{k\alpha} - \frac{\sin\left( k\beta \right)}{k\beta} - 1 \right]. 
\end{eqnarray}
Note that for $\gamma(k) > 0$, small values of $\alpha$ and $\sin\left( k\beta \right) < 0$ are required. From~\cite{DaCunha11}.

Initially glance, it may seem that Eq.~\eqref{App2} does not include a diffusive term. However, if we rewrite the first integral as~\cite{DaCunha11}
\begin{equation}
\label{Int} 
I(x) = a\displaystyle{\int_{0}^{L}}g_{\alpha}(x-x')u(x',t)dx' = a\displaystyle{\int_{x}^{L+x}}g_{\alpha}(y)u(y+x,t)dy = a\sum_{n=0}^{\infty} \lambda_n \frac{d^{2n}u(x,t)}{dx^{2n}},
\end{equation}
where we have used the Taylor expansion, that $g_\alpha(x)$ is an even function, the periodic boundary condition $u(x+L,t) = u(x,t)$, and
\begin{equation}
\label{IntCon} 
\lambda_n = \frac{1}{(2n)!}\int_0^L g(y) y^{2n}dy.
\end{equation}

Considering the first two terms in the expansion and $\beta \rightarrow 0$, we obtain the Fisher equation~\eqref{App3}, with $D = \frac{1}{6}a\alpha^2$. This was used to obtain the values of $\alpha$ and to describe successfully the diffusion of \emph{Escherichia coli} populations~\cite{DaCunha11}. Consequently, in the expansion Eq.~\eqref{Int}, the first term is the growth term, the second is the diffusive term, and the higher-order terms ($n > 1$) are the dispersive terms. Taking only the first two terms in the expansion and keeping the second integral, we get the formulation of Fuentes \textit{et al.}~\cite{Fuentes03,Fuentes04}. A considerable number of formulations may be regarded as particular cases of Eq.~\eqref{App2}, see, for example, Ref.~\cite{Aranda20-2}.

\begin{figure}[h]
\centering
\includegraphics[width=\linewidth]{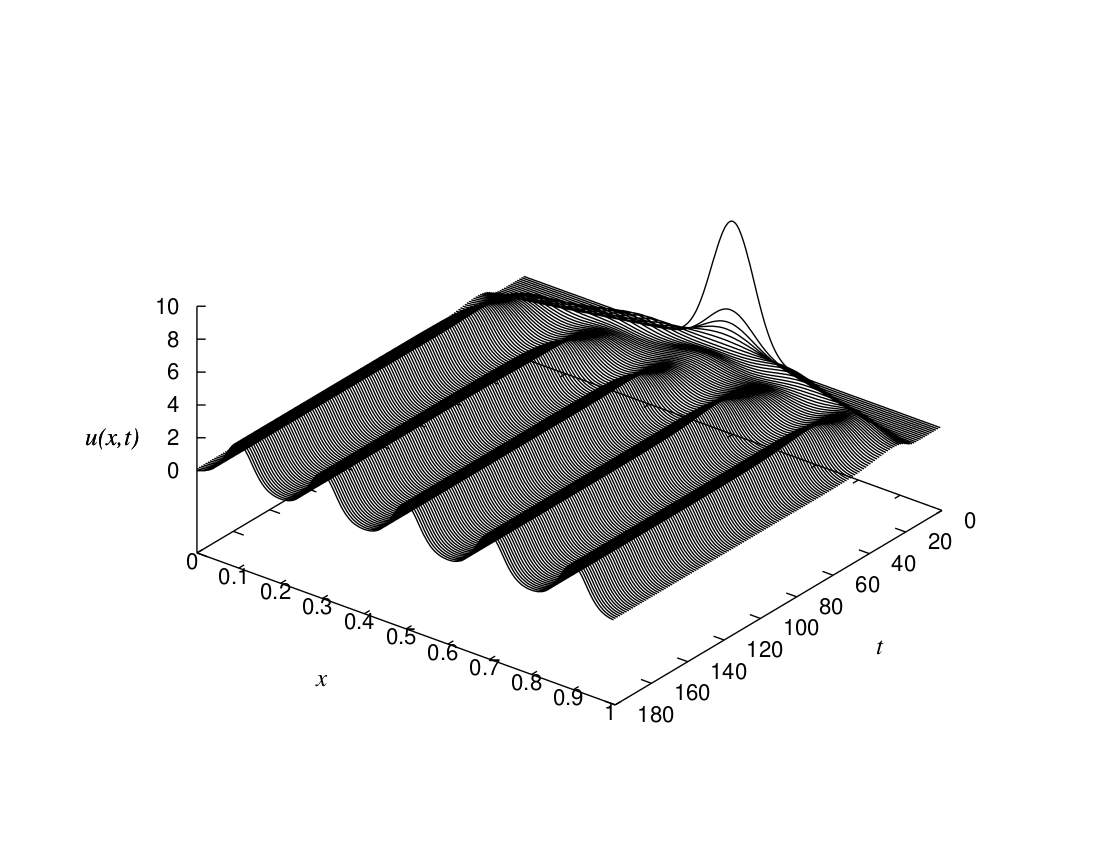}
\caption{Typical time and space evolution of the density $u(x,t)$. It starts as a Gaussian and, after a long time, $u(x,t)$ converges to {the} steady-state $u(x,t \rightarrow \infty) = U(x)$. {Adapted from various authors}~\cite{DaCunha11,Fuentes03,Fuentes04,DaCunha09}.}
\label{fig:density}
\end{figure}

In Figure~\ref{fig:density}, we show a typical curve of time and space evolution of the density $u(x,t)$. Here we use the values from Da Cunha \textit{et al.}~\cite{DaCunha11}. $\alpha$, $\beta$, and $x$ are given in units of $x$, and $u$ in units of $L^{-1}$. The growth and interaction rates are $a = b = 1$. The growth and competition length parameters are $\alpha = 0.009$ and $\beta = 0.15$, respectively. By setting the initial distribution as a Gaussian centered at $x = L/2$, the results are obtained from the solution of Eq.~\eqref{App2}. After a long time, it evolves to a steady-state solution $U(x)$, which means $u(x, t \rightarrow \infty) \rightarrow U(x)$. This type of evolution is found in many physical situations~\cite{Aranda20-1,Aranda20-2,Aranda21,DaCunha11,Fuentes03,Fuentes04,DaCunha09}.

Finally, we define a very important quantity
\begin{equation}
\rho = \int_0^L [U(x) - u_0]dx,
\end{equation}
i.e., considering $N_0 = u_0L$ as the number of individuals in a system with a uniform distribution, $\rho$ measures the deviation because of pattern formation, indicating that the breaking of translation symmetry creates a new order parameter.

\begin{figure}[h]
\centering
\includegraphics[width=\linewidth]{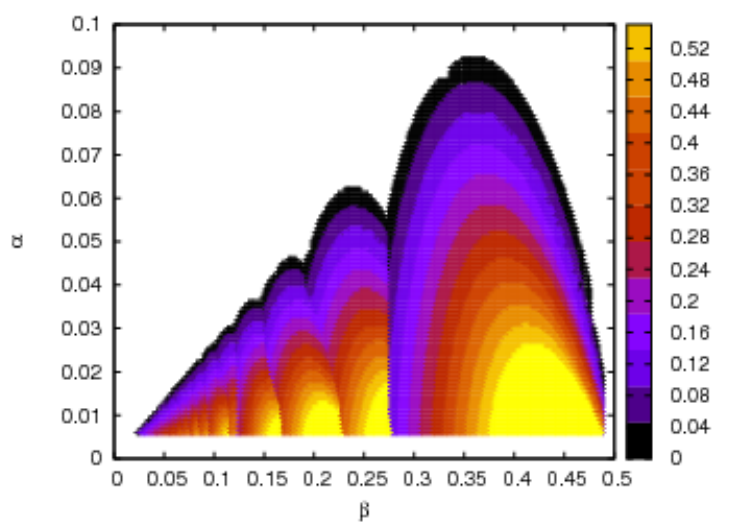}
\caption{The phase diagram in the $(\alpha,\beta)$ space. The parameter $\rho > 0$ showed by the colored column on the right, {indicates the addicional} population supported by the system {due to pattern formation}. From Barbosa \textit{et al.}~\cite{Barbosa17}.}
\label{fig:map}
\end{figure}

Thus, we show in Figure~\eqref{fig:map} the parameter $\rho = \rho(\alpha,\beta)$ in the space $(\alpha,\beta)$, representing the quantity that exceeds the population for a uniform density $N_0 = u_0L$. The results are from the solution of Eq.~\eqref{App2}, with $\rho$ in units of $N_0$, which is just a number. Regions that exhibit patterns are those with $\rho > 0$. Negative values of $\rho$ were not observed, i.e., $\rho \geq 0$, with the minimum value corresponding to uniform distributions. This shows that one reason nature creates patterns is to allow more individuals to exist in a given ecological niche.

In the above formulations, fluctuations are just small perturbations that may either die out, in which case $u(\vec{x},t \rightarrow \infty ) = u_0$, or grow, driving the system towards a pattern, i.e., a non-uniform density $u(\vec{x},t)$. We may include in population dynamics situations where the noise is strong~\cite{Fusco16}. However, we leave that for the next subsection, where we study growth phenomena. 

\subsection{Non-perturbative fluctuations}

Next, we will discuss a second kind of growth phenomenon that differs greatly from the previous one. Imagine a water reservoir of height $H$ and uniform lateral area. To determine how much water has been deposited in the reservoir over time $t$, it suffices to measure the time evolution of the height $h(t)$, $0 < h(t) < H$. In most physical systems, the growth process occurs when particles or aggregates of particles reach a surface via diffusion or injection or through some kind of deposition process of chemical or biological origin. For example, consider snow falling on the street as depicted in Figure~\ref{fig1}. This is similar to water being deposited in the reservoir, but now the interface is no longer uniform, and each point on the surface may have a height $h(\vec{x},t)$, where $\vec{x}$ is the position in the $d$-dimensional space and $t$ is the time. In this way, the height $h(\vec{x},t)$ and the position $\vec{x}$ form a $d+1$ dimensional space. This form of growth, where the interface is rough, has been intensively investigated over the past four decades. Some more examples include the growth of semiconductor thin films~\cite{Almeida14}, surface dissolution~\cite{Carrasco21}, corrosion or etching mechanisms~\cite{Mello01,Gomes19}, and fire propagation~\cite{Merikoski03}. The last three examples are of negative growth, i.e., $\langle \frac{\partial h(\vec{x},t)}{\partial t} \rangle < 0$. Thus, many physical systems in nature can exhibit one phase of growth, making it a prolific area of scientific research. In such systems, fluctuation-driven stochastic dynamics lead to a rough interface that separates the two media~\cite{Barabasi95,Edwards82,Kardar86,Hansen00}. Two quantities play an important role in growth dynamics: the average height, $\langle h(\vec{x},t) \rangle$, and the standard deviation
\begin{equation}
w(L,t) = \sigma_h = \left[ \langle h(\vec{x},t)^2 \rangle - \langle h(\vec{x},t) \rangle^2 \right]^{1/2}.
\label{wt}
\end{equation} 

The standard deviation provides a precise definition of the interface roughness, which is often referred to as roughness or surface width. Here, $\langle \cdots \rangle$ represents the space average, and $L$ is the system size. It should be mentioned that the dispersion in Eq.~\eqref{wt} is the main physical quantity to be obtained in the growth process. Indeed, in statistical physics, it is second only to diffusion, Eq.~\eqref{x20}, and many important phenomena have been associated with it.

\begin{figure}[!h]
\centering
\includegraphics[width=\linewidth]{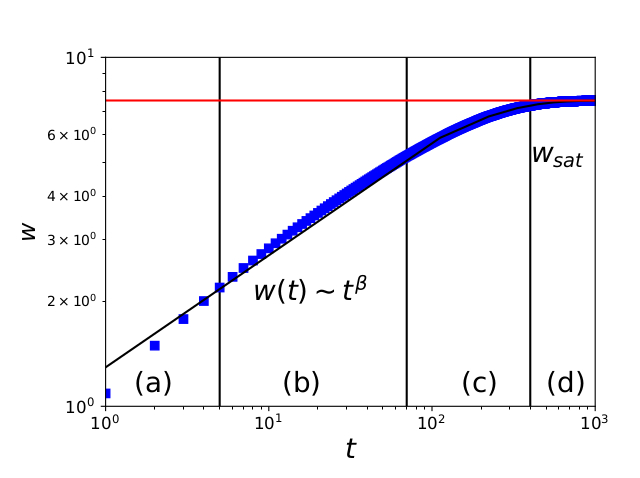}
\caption{Time evolution of the roughness, $w(t)$, as a function of time $t$ for the etching model {in $1+1$ dimensions}. The sample size is $L = 200$, and the time intervals are taken as $\Delta t = 1/L$, such that one site is reached on average per unit of time. The growth process exhibits different characteristics in each region: (a) $0 < t < t_0 \approx 5$, where correlations have not yet started, and the process is a random deposition with $w(t) \sim t^{1/2}$; (b) when correlations play a major role, $5 < t < 70$, the power-law region with $w(t) \sim t^{\beta}$; (c) the intermediate region $70 < t < 400$; (d) the saturation region with $w(t) \approx w_s$ for $t > 400$. From the figure, we obtain the characteristic time $t_\times = 167.0(2)$, indicating that saturation occurs for $t \gg t_\times$, as seen in region (d). From Gomes~\textit{et al.}~\cite{Gomes19}.}
\label{fig:W}
\end{figure}

The general characteristics of growth dynamics have been observed through analytical, experimental, and computational results~\cite{Barabasi95}. For many growth processes, the roughness $w(t)$ increases with time until it reaches a saturated roughness $w_s$, i.e., $w(t \rightarrow \infty) = w_s$. In Figure~\ref{fig:W}, we show the evolution of $w(t)$ as a function of $t$ for a typical growth process. We can summarize the time evolution of all regions as follows
\begin{equation}
\label{Sc1}
w(t,L) =
\begin{cases}
ct^{1/2}, &\text{if~~} t \ll t_0\\
ct^\beta, &\text{if~~} t_0 \ll t_\times\\
w_s \propto L^\alpha, &\text{if~~} t \gg t_\times.\\
\end{cases}
\end{equation}

Here $t_0$ is the time before correlations start, where we have pure diffusion, and $t_{\times} \propto L^z$. Note that the exponents $\alpha$ and $\beta$ here have no connection with the length parameters $\alpha$ and $\beta$ from the last section. Moreover, the dynamical exponents satisfy the general scaling relation~\cite{Family85,Rodrigues24}
\begin{equation}
\label{z}
z = \frac{\alpha}{\beta}.
\end{equation}

Different methods have been proposed to understand this rich phenomenon. Here, we will focus only on processes that saturate, i.e., those with a finite $w_s$, $w(t \rightarrow \infty) = w_s$. This occurs in a general class of growth phenomena. The attempt to describe the height evolution leads us to an equation of the form
\begin{equation}
\label{Gen2}
\dfrac{\partial h(\vec{x},t)}{\partial t} = \hat{G}h(\vec{x},t) + \xi(\vec{x},t),
\end{equation}
where $\hat{G}$ is a deterministic operator and the stochastic process is characterized by the noise, $\xi(\vec{x},t)$, which in its simplest form is assumed Gaussian and white
\begin{equation}
\label{xi}
\left\langle \xi(\vec{x},t) \xi(\vec{x'},t')\right\rangle = 2D\delta^{(d)}(\vec{x} - \vec{x'})\delta(t - t'),
\end{equation}
with $D$ representing the noise intensity. 

The deterministic part of the RHS of Eq.~\eqref{Gen2} must obey certain symmetry rules~\cite{Edwards82}, such as time translation and height invariance, as well as $h \rightarrow -h$ and $\vec{x} \rightarrow -\vec{x}$. These symmetry rules imply that only even-order derivatives are allowed, which brings us to some form of the Langevin equations. In the lowest order, the simplest field equation is~\cite{Barabasi95,Edwards82}
\begin{equation}
\label{EW}
\dfrac{\partial h(\vec{x},t)}{\partial t} = \nu \nabla^2 h(\vec{x},t) + \xi(\vec{x},t),
\end{equation}
known as the Edwards-Wilkinson equation (EWE)~\cite{Edwards82}. Here, the parameter $\nu$ is the surface tension, which tends to reduce the surface curvature. With this equation, we can explain certain forms of growth, such as deposition with relaxation~\cite{Barabasi95}. Note that if we exclude the noise, $D = 0$, Eq.~\eqref{EW} becomes the diffusion equation discussed in the Introduction section. The presence of space-inhomogeneous and time-dependent noise makes the process a more complex form of diffusion. For $1+1$ dimensions, the exponents are $\alpha = 1/2$, $\beta = 1/4$, and $z = 2$, in agreement with Eq.~\eqref{z}. 

A large number of systems, however, cannot be described by the EWE. Consider, for example, the roughness evolution in Figure~\ref{fig:W}, obtained from the etching model~\cite{Mello01} in $1+1$ dimensions. The etching model is described below. From that curve, we get $\beta = 0.330$, far from the Edwards-Wilkinson (EW) results. Consequently, new forms of growth equations have been proposed. A significant development was made when Kardar \textit{et al.}~\cite{Kardar86} introduced a new term into the EWE to obtain
\begin{equation}
\label{KPZ}
\dfrac{\partial h(\vec{x},t)}{\partial t} = \nu \nabla^2 h(\vec{x},t) + \dfrac{\lambda}{2}\left[\vec{\nabla}h(\vec{x},t)\right]^2 + \xi(\vec{x},t),
\end{equation}
where $\lambda$ is related to the tilt mechanism~\cite{Kardar86}. The nonlinear term arises when considering lateral growth. Note that for $\nu = 0$, we have the Burgers equation with noise. Now we have lost the symmetry $h \rightarrow -h$, leading to a new universality class. 

To summarize the basic information about KPZ equation, Eq.~\ref{KPZ}, we add~\cite{Kardar86}
\begin{equation}
\label{GI}
\alpha + z = 2.
\end{equation}

This is a consequence of Galilean invariance in the related Burgers equation~\cite{Forster97}.

\subsubsection{Universality}

In this way, the KPZ equation, Eq.~\eqref{KPZ}, is a general nonlinear stochastic differential equation that can characterize the growth dynamics of many different systems~\cite{Almeida14,Mello01,Merikoski03,Reis05,Almeida17,Alves16,Carrasco18,Nicoli13,Odor10,Takeuchi13,Rodrigues15}. Systems that exhibit the same exponents as KPZ belong to the KPZ universality class. As a consequence of this universality, most of these stochastic systems are interconnected. For instance, the single step (SS) model~\cite{Krug92,Krug97,Derrida98,Meakin86,Daryaei20}, which is connected with the asymmetric simple exclusion process~\cite{Derrida98}, the six-vertex model~\cite{Meakin86,Gwa92,Vega85}, and the kinetic Ising model~\cite{Meakin86,Plischke87}, are all of fundamental importance. It is noteworthy that quantum versions of the KPZ equation have been recently reported, which are connected with a Coulomb gas~\cite{Corwin18}, quantum entanglement growth dynamics with random time and space~\cite{Nahum17}, as well as in infinite temperature spin-spin correlations in the isotropic quantum Heisenberg spin-$1/2$ model~\cite{Ljubotina19,DeNardis19}, anisotropic quantum Heisenberg chain~\cite{Vainstein05}, and spin glass~\cite{Prykarpatski23}. 

In these processes, randomness is not a small perturbation; it is a strong component, and indeed $D$ is an important part of the theory. For example, the coupling parameter~\cite{Kardar86}
\begin{equation}
\label{g0}
g = \frac{D \lambda^2}{\nu^3},
\end{equation}
connects all KPZ parameters and is a fundamental quantity for any renormalization process.

Despite all efforts, finding an analytical or even a numerical solution to the KPZ equation~\eqref{KPZ} is challenging~\cite{Dasgupta96,Dasgupta97,Torres18,Wio10a,Wio17,Rodriguez19}, and we are still far from a satisfactory theory for the KPZ equation. This makes it one of the most difficult and exciting problems in modern mathematical physics~\cite{Bertine97,Baik99,Prahofer00,Dotsenko10,Calabrese10,Amir11,Sasamoto10,Doussal16,Hairer13}, and probably one of the most important problems in non-equilibrium statistical physics. The outstanding works of Pr{\"a}hofer and Spohn~\cite{Prahofer00} and Johansson~\cite{Johansson00} opened the possibility of an exact solution for the distributions of height fluctuations $f(h, t)$ in the KPZ equation for $1 + 1$ dimensions~\cite{Calabrese10,Amir11,Sasamoto10,Doussal16}.

To define the fields of action in the investigation of KPZ dynamics, we have the following main scenarios:
\begin{enumerate}
\item[a.] Describing experiments using KPZ dynamics;
\item[b.] Matching a given model into KPZ;
\item[c.] Determining the exponents and height distribution;
\item[d.] Matching a given KPZ model into another KPZ model.
\end{enumerate}

We define a KPZ model as a model that belongs to the KPZ universality class. In our current stage of knowledge, only in $1+1$ dimensions have these questions received precise answers.

\subsection{Experiments described by Kardar-Parisi-Zhang dynamics}

Here, we briefly present some experiments well described by the KPZ approach. 

\begin{figure}[!h]
\centering
\includegraphics[width=\linewidth]{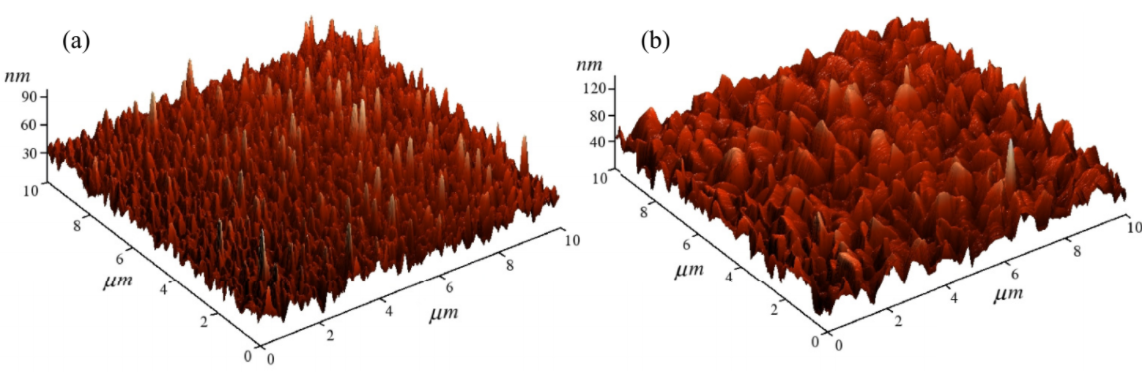}
\caption{The surface morphology growth for polycrystalline CdTe/Si(100) films. AFM images were taken for a sample with dimensions of $10 \, \rm \mu m \times 10 \, \rm \mu m$ at deposition times $(a)$ $t = 60$ min; $(b)$ $t = 240$ min. From Almeida \textit{et al.}~\cite{Almeida14}.}
\label{fig_Almeida}
\end{figure}

The first is a typical growth experiment, where polycrystalline CdTe/Si(100) films grow by hot-wall epitaxy~\cite{Almeida14}. In Figure~\ref{fig_Almeida}, we show the evolution of morphology at different times during growth. The roughness of the surface becomes clear. From this, the exponent $z = 1.61(5)$ was obtained. Note that while the sample sizes are measured in $\mu m$, the roughness displayed on the $z$ axis is in $nm$. 

The experimental scenario for exploring the mutation and propagation of bacteria is very rich with many possibilities. Thus, it is important to mention some interesting results~\cite{Fusco16} in expanding populations revealed by spatial Luria-Delbrück experiments. There, the roughness front was measured, and it was reported that for a flat front, they observed $z = 2$ for both $1+1$ and $2+1$ dimensions, corresponding to the EW universality class. For a rough front, they observed $z = 3/2$ for $1+1$ and $z = 1.61$ for $2+1$ dimensions, corresponding to the KPZ universality class, in agreement with~\cite{GomesFilho21b}.

\subsection{Discrete growth models}

Obtaining a solution for the KPZ equation in $d+1$ dimensions is a significant challenge, both analytically and numerically. Indeed, it is difficult even for $d=1$. On the other hand, there are several discrete models that operate by simple rules, which can be easily implemented in a computer code. Such models are sometimes referred to as cellular automata models~\cite{Barabasi95,Almeida14,Mello01,Reis05,Almeida17,Alves16,Carrasco18,Odor10,Rodrigues15,Mello15}. Matching some models to the KPZ universality class is part of scenario $c$.

As stated above, simulations of a cellular automaton are simpler and more precise than the numerical solution of the KPZ equation itself. Here, we show two types of cellular automaton models that belong to the KPZ universality class. 

\subsubsection{The single step model}

The single step (SS) model has gained significant importance over the years~\cite{Krug92,Krug97,Derrida98,Meakin86,Daryaei20} for two reasons: first, because it was proven to be a KPZ model (scenario $c$), and second, because of its connection with other models, such as the asymmetric simple exclusion process~\cite{Derrida98}, the six-vertex model~\cite{Meakin86,Gwa92,Vega85}, and the kinetic Ising model~\cite{Meakin86,Plischke87}, making it a good example of scenario $d$. 

The SS model is defined such that the height difference between two neighboring heights $\eta = h_i - h_j$ is just $\eta = \pm 1$. Consequently, it is easily associated with the Ising model. For example, in $1+1$ dimensions, the initial conditions for the height of site $i$ take the form $h_i(0) = (1 + (-1)^i)/2$. Now, consider a hypercube of side $L$ and volume $V = L^d$. We select a site $i$, compare its height with that of its neighbors, and apply the following rules:\begin{enumerate}\item At time $t$, randomly choose a site $i \in V$;\item If $h_i(t)$ is a minimum, do $h_i(t+\Delta t) = h_i(t) + 2$, with probability $p$;\item If $h_i(t)$ is a maximum, do $h_i(t+\Delta t) = h_i(t) - 2$, with probability $q$.\end{enumerate}

Using these rules, we can generate the dynamics of the SS model in $d+1$ dimensions. For $1+1$ dimensions, its properties have been well studied~\cite{Krug92,Krug97,Derrida98}, and we can obtain exact analytical results, such as\begin{equation}\label{lamb0}\lambda = p - q.\end{equation}

Note that, for $p = q$, it becomes the EW model. By changing the probabilities, we can get a lot of relevant information~\cite{Daryaei20}.

\begin{figure}[!h]
\centering
\includegraphics[width=0.8\linewidth]{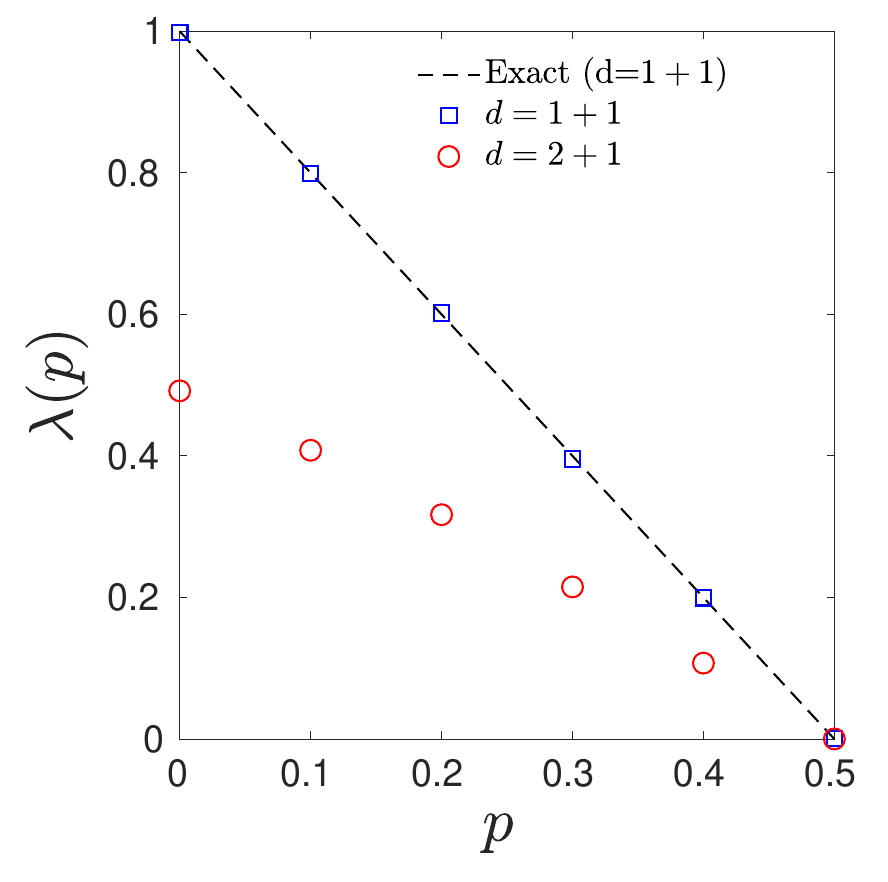}
\caption{The parameter $\lambda$ as a function of the probability $p$ for the SS model. For $1+1$ dimensions, we have the exact result $\lambda = p - q$; for $2+1$ dimensions, $\lambda \propto (p - q)^\delta$, with $\delta = 0.87(4)$. For $p = q$, we have the Edwards-Wilkinson model. From Daryaei~\cite{Daryaei20}, with permission from PRE.}
\label{fig_lamb}
\end{figure}

\begin{figure}[!h]
\centering
\includegraphics[width=\linewidth]{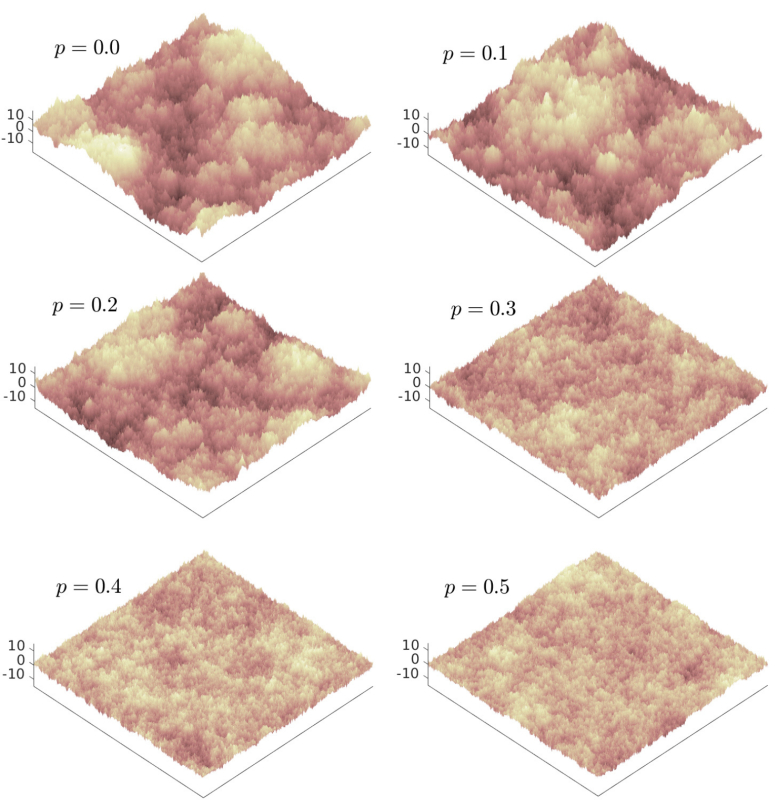}
\caption{The surface morphology growth for the SS model at various $p$ values with a lattice size of $1024 \times 1024$. From Daryaei~\cite{Daryaei20}.}
\label{fig_surf}
\end{figure}

In Figure~\ref{fig_lamb}, we show $\lambda$ as a function of the probability $p$. We exhibit the exact result for $1+1$ dimensions and the simulations for $2+1$ dimensions, where $p + q = 1$. From~\cite{Daryaei20}. In Figure~\ref{fig_surf}, we show snapshots of a typical surface morphology grown by SS in $2+1$ dimensions, at the steady-state regime, with $L = 1024$ and $p$ ranging from $p = 0$ (the ``best'' KPZ regime) to $p = 0.5$, i.e., $\lambda = 0$, which characterizes the EW regime.

\subsubsection{The etching model}

The etching model~\cite{Mello01,Gomes19,Alves16,Rodrigues15} is a stochastic cellular automaton that simulates surface erosion because of the action of acid. This model is designed to be as simple as possible while capturing the essential physics, considering that the probability of removing a cell is proportional to the number of exposed faces of the cell (an approximation of the etching process). First, we randomly select a site $i$ with a certain height $h_i$, and then we compare it with one of its nearest neighbors $j$. This is similar to the SS model. Now the rules change: If $h_j > h_i$, it is reduced to the same height as $h_i$, which means that the surface height decreases at each step. The main algorithm steps are summarized as follows:\begin{enumerate}\item At time $t$, randomly choose a site $i \in V$;\item If $h_j(t) > h_i(t)$, do $h_j(t+\Delta t) = h_i(t)$;\item Do $h_i(t+\Delta t) = h_i(t) - 1$.\end{enumerate}

This stochastic cellular automaton has the advantage of allowing us to understand a corrosion process while also studying the KPZ growth process. 

Note that rule 1 is common to all these models and is equivalent to the random term $\xi(\vec{x},t)$ in the KPZ equation~(\ref{KPZ}). The interactions between neighbors are equivalent to diffusive and nonlinear terms (lateral growth). However, both $\nu$ and $\lambda$ vary in intensity for each model.

We now apply the above rules to study the etching mechanics in a system in $1+1$ dimensions with size $L = 200$. To illustrate the main characteristics of a standard surface width evolution, we present a typical curve of roughness $w(t)$ as a function of time $t$ in Figure~\ref{fig:W}, from reference~\cite{Gomes19}. The results were averaged over the sites and over $1 \times 10^7$ different simulations. They also considered systems with periodic boundary conditions. The time evolution of the roughness dynamics, $w(t)$, presents four different regimes: $(a)$ at the beginning of the process (without correlations), $t < t_0 \approx 5$, the mechanism is classified as random deposition, $w(t) \approx t^{1/2}$, that is, basically normal diffusion; $(b)$ the second region, $5 < t < 70$, where the system has had enough time for correlations to reach distant points of the lattice, and the roughness is described by a power-law dependence, $w(t) \approx t^{\beta}$; for $t$ between $70$ and $400$, the system passes through an intermediate region $(c)$; and finally, in the last region $(d)$, for times greater than $400$, the system reaches the saturation region $w(t) \approx w_s$. Three major results were obtained from the figure: the characteristic time $t_\times = 167.0(2)$; the saturation roughness $w_s = 7.5228(6)$ for $t \gg t_\times$; and the exponent $\beta$ was found to be $0.334(3)$, which is very close to the exact KPZ value ($1/3$).

As an example of obtaining the KPZ equation from a discrete model, Gomes \textit{et al.}~\cite{Gomes19} recently developed a procedure that was successfully applied to the above etching model.

To obtain the KPZ equation from the etching model, we follow this procedure: First, we define $P(\eta,t)$ as the probability distribution that if at time $t$ a site $i$ has height $h_i$, its neighbor must have height $h_i + \eta$, i.e., $\eta = h_i - h_j$ is the height difference between two neighboring sites.

Finally, we apply the rules of the etching model to obtain the KPZ equation with the coefficients as in~\cite{Gomes19}\begin{equation}\label{D00}D = \omega_0d,\end{equation}where $\omega_0 = 1/\Delta t$ is the frequency of the noise, which drives all processes and appears in all coefficients. The linear dependence on $d$ is similar to diffusion models~\cite{Oliveira19,Santos19}. Note that for the KPZ type of model, $P(\eta,t)$ stabilizes quickly and can be considered time-independent for all regions of interest~\cite{Gomes19}; consequently, $P(\eta,t) = P(\eta)$. Then, we also have\begin{equation}\label{nu}\nu = \frac{\omega_0d}{2} P(0) = \frac{P(0)}{2}D,\end{equation}and\begin{equation}\label{lambWal}\lambda = 2\omega_0d\frac{dP(\eta=0^+)}{d \eta} = 2D\left(\frac{dP(\eta)}{d \eta}\right)_{\eta=0^+}.\end{equation}

Thus, Eqs.~(\ref{D00}) to~(\ref{lambWal}) establish a very important relationship between the KPZ parameters $\lambda$ and $\nu$ and $P(\eta)$, showing that both KPZ parameters depend on lateral growth and are more than just phenomenological parameters~\cite{Gomes19}.

It should be noted that the increase in $\lambda$ with dimension is a strong argument supporting Eq.~\eqref{GI}, as this shows that $\lambda$ drives the other parameters. This method may be useful for other cellular automata models.

Therefore, this was an example of a cellular automaton model for growth and a recent method used to obtain the KPZ equation from it. Moreover, it allows the calculation of the coefficients $\nu$, $D$, and $\lambda$. The values of these coefficients change with the model and dimensions, while the exponents $\alpha$, $\beta$, and $z$ are universal and depend only on the dimensions $d$.

For the etching cellular automaton, the coupling constant $g$ becomes
\begin{equation}
\label{g}
g = \frac{D\lambda^2}{\nu^3} = \frac{32}{P(0)^3}\left(\frac{dP(\eta)}{d \eta}\right)^2_{\eta=0} = \frac{32}{P(0)}\left(\frac{d \ln(P(\eta))}{d \eta}\right)^2_{\eta=0}.
\end{equation}

This is the ultimate goal of statistical physics: to express a given quantity in terms of probabilities~\cite{Gomes19}. Unfortunately, this method applies only to the etching model. It would be very interesting if a similar approach could be developed for other models.

Different methods and models may reveal new features and the depth and beauty of KPZ dynamics. However, the major limitation of this method is that we do not have an analytical expression for $P(\eta)$. The search for such an expression would represent a significant advancement in the field.

\subsubsection{Universal properties of growth models}

Universality extends beyond the set of critical exponents, encompassing height distributions and spatial and temporal correlations. In recent years, it has been shown that these other universal quantities depend on the interface geometry (see~\cite{Kriecherbauer10,Halpin-Healy15} for recent reviews), where circular and linear interfaces exhibit different statistics. Later, simulations of flat surfaces that expanded over time indicated that the dynamics of the system size were the true cause of the subdivision. Note that while a linear interface has a fixed lateral size, the perimeter of a circular one grows with its radius.

\begin{figure}[!h]
\centering
\includegraphics[scale=1]{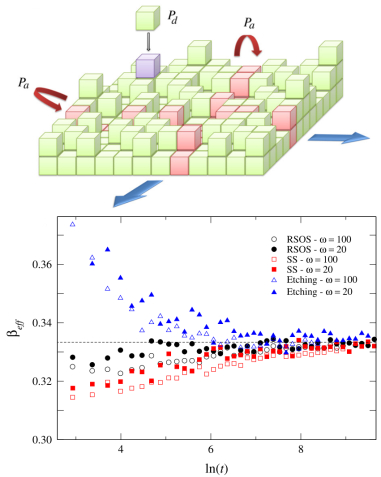}
\caption{(Top) Illustration of the substrate enlargement and deposition for $2 + 1$ dimensions. (Bottom) The evolution of the exponent $\beta$ as a function of $t$ for $1+1$ dimensions. From Carrasco \textit{et. al.}~\cite{Carrasco14}.}\label{fig_enlarging}
\end{figure}

For instance, Figure~\ref{fig_enlarging} illustrates the lateral enlargement for simulations in $2+1$ dimensions. At each time step, there is a chance of particle deposition with a probability $P_d = \frac{N}{N + \omega d_s}$, and a probability $P_a = \frac{\omega d_s}{N + \omega d_s}$ for the system to enlarge through the duplication of a row or a column (see~\cite{Carrasco14} for more details). Here $N$ is the total number of sites, $d_s = d + 1$, i.e., the ``full'' volume, and $\omega$ is the enlargement rate, i.e., the size evolves as $\left\langle L(t)\right\rangle = L_0 + \omega t$. Even with no curvature, adding these lateral duplications was enough for the system to display the statistics of the curved KPZ subclass. The irrelevance of the geometry in the subdivision was confirmed when experiments with curved interfaces that shrank over time showed the statistics of the flat KPZ subclass~(\cite{Fukai17}, further discussed in~\cite{Carrasco18,Carrasco19}). In the bottom part of Figure~\ref{fig_enlarging}, we show the values of $\beta$ as a function of $t$ for some models in $1+1$ dimensions. Here $\beta(t) = \frac{d \ln {[W(t)]}}{dt}$. We observe convergence to $\beta = 1/3$, as expected for the KPZ universality class in $1+1$ dimensions.

\section{Restoring the fluctuation-dissipation theorem through a new emergent fractal dimension} \label{sec:fractal} 

As previously discussed, the FDT fails in many situations, primarily due to violations of ergodicity (see also Section \ref{sec:Beyond}). In this section, we explore the possibility of recovering the FDT by transitioning from Euclidean to fractal space. We focus on two significant examples: growth phenomena and phase transitions.

\subsection{Fluctuation-dissipation theorem for the Kardar-Parisi-Zhang equation} \label{Sec_FDTKPZ}

A closer examination of Eqs.~(\ref{nu}) and~(\ref{lambWal}) reveals not only a dependence on $P(\eta)$ but also a clear influence of the noise $D$ on the parameters $\lambda$ and $\nu$. This indicates that the fields on the right-hand side of Eq.~\eqref{KPZ} originate from the same source. This represents a form of the FDT associated with the parameters of a solid-on-solid model~\cite{Gomes19}, serving as an initial step towards understanding the role of the FDT in growth phenomena.

For the KPZ equation, the FDT is valid only in $1+1$ dimensions~\cite{Kardar86,Rodriguez19}. The primary issue lies in Langevin's equation, where both fluctuation and dissipation are components of the same force because of the thermal bath—an observation not clearly evident in the growth phenomena. A recent article~\cite{GomesFilho21} demonstrated that the saturation mechanism, i.e., $w(t\rightarrow \infty)=w_s$, is analogous to the thermalization mechanism in Langevin's equation.

For both the EW and KPZ equations, it is observed that noise increases $h$ and $w(t)$ linearly with time, while the term $\nu \nabla^2 h(\vec{x},t)$, which acts against roughness, decreases it. Roughness increases with time until it reaches saturation, which, for $1+1$-dimensional systems, is given by the exact results
\begin{equation}
\label{Ws}
w_s=\sqrt{\frac{cD}{\nu}L},
\end{equation}
where $c=1/(4 \pi^2 )$ for the EW model~\cite{Edwards82} and $c=1/24$ for the SS model~\cite{Krug92,Krug97}. Although there is no equipartition theorem for growth, we have $w^2(t \rightarrow \infty) = w_s^2$, analogous to $\left\langle v^2(t \rightarrow \infty) \right\rangle = k_BT/m$. By substituting $D$ with the above value, Eq.~\eqref{xi} becomes
\begin{equation}
\label{GFDT}
\left\langle \xi({x},t) \xi({x'},t')\right\rangle = 2b\nu w_s^2\delta({x}-{x'})\delta(t-t'),
\end{equation}
with $b=1/(cL)$. Therefore, the parameter $D$ in Eq.~\eqref{xi} is not only the noise intensity but is also related to $\nu$, leading to a FDT. Note the similarity between Eq.~\eqref{GFDT} and Eq.~\eqref{FDT0}. Additionally, since the noise and surface tension in the EWE originate from the flux, their separation is artificial. Consequently, this discussion restores the lost link~\cite{GomesFilho21,Anjos21}.

Noting that the growth velocity $v_g$ is given by
 \begin{equation}
 \label{vg}
v_g=\langle	\partial_{t}h(\mathbf{x},t) \rangle =  \frac{\lambda}{2} \langle\left( \nabla h \right)^2 \rangle,
\end{equation}
$v_ g$ measured in deposition layers is a constant. We can rewrite $ (\lambda/2) \left( \nabla h \right)^2 = v_g + (\lambda/2) \left( \nabla h \right)^2 - v_g=v_g+\xi_v$, where $\xi_v=(\lambda/2) \left( \nabla h \right)^2 - v_g$ is the fluctuation in the growth velocity, i.e. a noise. Thus, the total noise $\eta({x},t)=\xi({x},t)+\xi_v({x},t)$ has zero mean and satisfies:
\begin{equation}
  \left\langle \eta(\vec{x},t) \eta(\vec{x'},t')\right\rangle = 2D_{eff}(t)\delta^{({d_n(t)})}(\vec{x}-\vec{x'})\delta(t-t').
\end{equation}

As the roughness of the interface increases over time, the intensity of this new noise is affected by the interface's fractality. As a result, the emergent noise follows a new FDT with a fractal dimension $d_n(t)$, which evolves over time until the saturation is reached at $w(t \gg t_x) = w_s$, i.e. $d_n(t \gg t_x) = d_n$ and {$D_{eff}(t \gg t_x)=D_{eff}^*$}.

Now Eq.~\eqref{KPZ} becomes
\begin{equation}
\label{KPZfr}
\dfrac{\partial h(\vec{x},t)}{\partial t} = \nu \nabla^2 h(\vec{x},t) + v_g + \eta(\vec{x},t).
\end{equation}
Thus, the balance is established between the Laplacian and the new noise terms, with no interference 
 from $v_g$ (constant). Moreover, from a proper dimensional analysis, {$[w_s] = [D_{eff}^*L/\nu]^\alpha$}, one can obtain a relation between the exponent $\alpha$ and the fractal dimension of the noise $d_n$ as:~\cite{GomesFilho24}
\begin{equation}
\alpha=\frac{1}{{d}_n+1},
\label{alf}
\end{equation}
with $d-1 \leq {d}_n \leq d$. In the case of $d=1$, the only possibility is ${d}_n=1$, resulting in the exact value~\cite{GomesFilho24} $\alpha=1/2$. Conversely, we can write
\begin{equation}
D_{eff}^* \propto (w_s/L^\alpha)^{1/\alpha}\nu,
\label{FDTS}
\end{equation}
Noting from Eq.~\eqref{Sc1} that $w_s/L^\alpha$ is a constant, Eq.~\eqref{FDTS} is a perfect form of FDT, that is, a linear relationship between the fluctuation {$D_{eff}^*$} and the dissipation $\nu$. Thus, Eq.~\eqref{FDTS} recovers the FDT for KPZ~\cite{GomesFilho24}.

\subsubsection{The search for Kardar-Parisi-Zhang exponents}

The search for KPZ exponents has been ongoing for over three decades. For $1+1$ dimensions, the Renormalization Group (RG) method~\cite{Kardar86} yields $\alpha=1/2$, $\beta=1/3$, and $z=3/2$. However, the RG method does not provide accurate results for $d \geq 2$. Since Eq.~\eqref{z} and Eq.~\eqref{GI} relate the three exponents, determining one of them allows us to infer the others. We just need one more equation. Eq.~\eqref{alf} provides a new relation, but it introduces a new unknown, $d_n$. Therefore, in the following subsections, we focus on determining the value of $\alpha$.

Based on empirical data, Wolf and Keztesz~\cite{Wolf87} proposed
\begin{equation}
\label{WK}
\alpha(d)= \frac{1}{d+1},
\end{equation}
while Kim and Kosterlitz~\cite{Kim89} proposed
\begin{equation}
\label{KK}
\alpha(d)= \frac{2}{d+3}.
\end{equation}
With the development of better computational tools and machines, simulations have shown values that diverge from both proposals. Lassig~\cite{Lassig98} proposed a field theoretical method yielding $\alpha(2)=2/5$ and $\alpha(3)=2/7$, which are again far from the simulation results. Up to now, theoretical attempts can be summarized as follows:
\begin{enumerate}
\item Scaling fails for all dimensions.
\item RG gives exact results only for $1+1$ dimensions~\cite{Kardar86}.
\item Field theoretical methods yield imprecise exponents~\cite{Lassig98,Sun94}.
\end{enumerate}

The RG approach does not provide precise exponents for $d+1$ dimensions for $d \geq 2$. However, some calculations using RG produce useful results. For example, Canet \textit{et al.}~\cite{Canet10,Canet11} used non-perturbative RG to obtain the only complete analytical approach yielding a qualitatively correct phase/flow diagram to date.

\subsubsection{Kardar-Parisi-Zhang exponents in $d+1$ dimensions}

The attempt to obtain the exact exponents for the KPZ equation, as discussed above, is one of the major objectives in this area. However, this goal has not yet been achieved. Since $2+1$ dimensions correspond to our three-dimensional space, this is physically the most important dimension. Therefore, we highlight some recent proposals for the KPZ exponents~\cite{GomesFilho21b}.

The key idea is that while the space dimension is $\tilde{D}=d+1$, the interface dimension $D_i$ where the growth occurs is fractal~\cite{Anjos21}, given by $D_i=d_f+1$, where $d_f$ is the interface fractal dimension. The presence of fractals in nature is undeniable, and accounting for these effects is increasingly necessary~\cite{Barabasi95,Avnir83,Bunde13,Luis22,Luis23}. The fractal nature of the interface is evident in Figure~\ref{fig_surf} and the experimental results in Figure~\ref{fig_Almeida}. The roughness exponent $\alpha$ is connected to the interface fractal dimension $d_f$ by~\cite{Barabasi95}
\begin{equation}
\label{dfe}
\alpha = 
\begin{cases}
2-d_f, &\text{ for } d =1, \\
d-d_f, &\text{ for } d \geq 2,
\end{cases}
\end{equation}
 Note that a second relation connecting $\alpha$ and $d_f$ is needed as well to obtain ($d_f,\alpha,\beta,z$). Now for $2+1$ dimensions, a good approximation is ${d}_n=d_f$, which allows one to obtain the KPZ exponents in $2+1$ dimensions~\cite{GomesFilho24}
\begin{equation}
\label{alfn}
z=d_f = \frac{1+\sqrt{5}}{2}, \hspace{0.7cm} \alpha = \frac{3-\sqrt{5}}{2}, \hspace{0.7cm} \beta = \sqrt{5}-2.
\end{equation}

These exponents are consistent with various experimental results. For example, accurate experiments measuring the exponent $z$ have reported values such as $1.6(2)$~\cite{Orrillo17}, $1.6(1)$~\cite{Ojeda00}, $1.61(5)$~\cite{Almeida14}, and $1.61(2)$~\cite{Fusco16}, which agree with the value $z = d_f = \frac{1+\sqrt{5}}{2} = 1.61803...$. Recent simulation results~\cite{Luis22,Luis23} further support these findings, having obtained similar values of $d_f$ using different methods.

The exponents also show good agreement with simulations. For instance, the average over $20$ different methods and models~\cite{GomesFilho21b} gives $\overline{\alpha}=0.3828 \pm 0.0037$, which aligns well with $\alpha = \frac{3-\sqrt{5}}{2} = 0.381966011$. Although these results are approximate, there is currently no exact solution accepted by the scientific community. Consequently, the search for the exact KPZ exponents remains ongoing.

An exact relation $d_n=f(d_f)$ would solve the problem. While this may not initially seem like progress, in Section \ref{sec:phase}, we show that the similar ideas lead to an exact solution for the second-order equilibrium phase transition.

\subsubsection{No upper critical dimension for Kardar-Parisi-Zhang}

The study of models in dimensions higher than $2+1$ has significant theoretical importance, including the investigation of the existence of an Upper Critical Dimension (UCD) where the exponents no longer change. This concept is well-known in the case of the Ising model, which has a UCD of 4. Simulations~\cite{Odor10, Rodrigues15, Alves14} suggest that if a UCD exists for the KPZ universality class, it must be for $d > 6$. However, Gomes-Filho \textit{et al.}~\cite{GomesFilho24} proved the non-existence of a UCD. For $d > 2$, Eq.~\eqref{alf} with $d-1 \leq {d}_n \leq d$ imposes the limit~\cite{GomesFilho24}
\begin{equation}
\frac{1}{d+1} \leq \alpha = \frac{1}{d_n+1} \leq \frac{1}{d},
\end{equation}
which means that as $d$ changes, the allowed values for the exponents also change. Consequently, there is no UCD for KPZ.

As mentioned above, the wide application of the methods developed for KPZ dynamics and the large number of associated models, along with the new wave of quantum phenomena recently mapped~\cite{Corwin18,Nahum17,Ljubotina19,DeNardis19} into KPZ, makes it one of the biggest problems in statistical physics. This underscores how central this question is, and how far we are from exhausting the topic.

\section{Recovering the fluctuation-dissipation theorem in phase transitions: Exact results} \label{sec:phase}

It is well known that near a phase transition, the FDT fails due to the inadequacy of mean field theory. The primary idea in this work is to associate thermodynamic responses with fluctuations in an appropriate fractal space, thereby recovering the FDT near a transition. This approach involves examining response functions, as exemplified by Eqs.~(\ref{SigE}), (\ref{SigV}), (\ref{SigM}), and (\ref{Sigr}), which represent the relationship between fluctuations and their corresponding thermodynamic responses. Here, we demonstrate both the failure and recovery of the FDT near phase transitions. The failure is clear when using traditional mean field approaches, which do not accommodate the critical fluctuations' spatial correlations. However, by employing a fractal space framework, we can recover the FDT, as the fractal geometry more accurately represents the system's underlying structure near critical points.

\subsection{The fluctuation-dissipation theorem failure}

In equilibrium phase transitions, we can define a static correlation function
\begin{equation}
  G(r)=\langle \chi(\vec{r}+\vec{y})\chi(\vec{y})\rangle,
\end{equation}
where $\chi(\vec{r})$ represents a fluctuating field, such as liquid density fluctuation, magnetization fluctuation, or charge density fluctuation. The average $\langle \ldots \rangle$ is taken over space, which can be either a lattice or continuous. In the continuous limit, we can derive the geometric dependence of $G(r)$ as~\cite{Lima24}
\begin{equation}
\left(-\nabla^2 +\kappa^2\right)G(r)=\delta^d(r),
\label{G}
\end{equation}
with the solution $G(r) \propto r^{2-d} \exp(-r/\rho)$, where $d$ is the Euclidean space dimension and $\rho=\kappa^{-1}$ is the correlation length.

However, the mean field solution breaks down at the transition, necessitating an empirical correction through the introduction of the Fisher exponent $\eta$~\cite{Fisher64},
\begin{equation}
\label{G2}
G(r) \propto
\begin{cases}
r^{2-d} \exp(-r/\rho), & \text{if } r > \rho, \\
r^{2-d-\eta}, & \text{if } r \ll \rho.
\end{cases}
\end{equation}

Here we observe a mean field failure, which is a failure of the FDT, which incorrectly predicts $\eta=0$, which is evidently inaccurate as $\eta$ depends on the dimension and the universality class of the model. For instance, the $2d$ Ising model has $\eta=1/4$. For temperatures $T$ close to the critical temperature $T_c$, the correlation length diverges as
\begin{equation}
\label{rhodivergence}
\rho \propto |T-T_c|^{-\nu}.
\end{equation}
\subsection{Scaling and fractals}

The critical exponent $\nu$ is related to the critical specific heat exponent $\alpha$ through the hyperscaling relation~\cite{Cardy96}
\begin{equation}
\label{alpha}
\alpha=2-d\nu,
\end{equation}
thus associating a thermodynamic variable with the divergence of the correlation length. Additionally, we have the Rushbrooke equality,
\begin{equation}
\label{rush}
\alpha + 2\beta + \gamma = 2,
\end{equation}
and the Fisher relation~\cite{Fisher64},
\begin{equation}
\label{gamma}
\gamma = (2-\eta)\nu,
\end{equation}
where $\beta$ is the order parameter exponent and $\gamma$ is the susceptibility exponent~\cite{Salinas01,Cardy96}. These relationships are derived using the concept of scale invariance.

Fractals are statistically self-similar geometric objects characterized by non-integer dimensionalities~\cite{Feder88}. Interpreting the scaling behavior observed at critical points as manifestations of fractal geometries is a well-researched field~\cite{Devakul19,Kroger00,Suzuki8}. In this way, fractals are directly related to scaling and multiscaling~\cite{Aravena24}.

The relation between the fractal dimension of the ordered phase $d_l$, i.e. its largest cluster, and $\beta$ was first proposed by Suzuki~\cite{Suzuki8}
\begin{equation}
\label{dl}
d_l = d - \frac{\beta}{\nu}.
\end{equation}

If the critical point is the percolation threshold, the fractal structure is represented by the infinite percolating cluster at the transition~\cite{Cruz23,Grimmett06}. In general, it is associated with the largest ordered cluster at the critical point~\cite{Kroger00}.

\subsection{Recovering the fluctuation-dissipation theorem}

Motivated by the recovery of the FDT in KPZ~\cite{GomesFilho24} and by the association of critical exponents with fractal dimensions~\cite{Kroger00,Suzuki8}, Lima \textit{et al.}~\cite{Lima24,Lima25} considered that Eq.~\eqref{G} at the transition $T \rightarrow T_c$, where $\kappa = 1/\rho \rightarrow 0$, should be formulated in fractal space. They used the Riesz fractional derivative of order $\zeta$, $0 < \zeta < 1$, associated with the fractal dimension $d_f$, $d-1 < d_f < d$, to obtain
\begin{equation}
\label{G3}
(-\nabla^2)^\zeta G(r) = \delta^{d_f}(r),
\end{equation}
to reflect the dynamics being restricted to a fractal structure characterized by the correlation fractal dimension $d_f$. {The solution of Eq.~\eqref{G3}, with the imposition of Eq.~\eqref{G2} yields~\cite{Lima24}}
\begin{equation}
\label{dff}
\eta = d - d_f = 1 - \zeta.
\end{equation}

Thus, the Fisher exponent $\eta$ in the correlation function $G(r)$ represents the deviation of the correlation fractal dimension from the integer dimension. Both $\zeta$ and $d_f$ are within the previously defined limits $0 < \zeta < 1$ and $d-1 \leq d_f \leq d$. Note that the Fisher exponent appears naturally, without the need for an \textit{ad hoc} assumption. They also found a connection between $d_l$ and $d_f$ as
\begin{equation}
\label{dl2}
d_f = 2(d_l-1).
\end{equation}

Note that both Eq.~\eqref{dff} and Eq.~\eqref{dl2} are exact results. As shown in Table \ref{Table1}, where $d_l$ and $d_f$ are listed for the Ising model in 2, 3, and 4 dimensions along with the relevant critical exponents, this equation holds true for all dimensions shown. Note also that as $d$ increases, $d_f$ approaches $d$, reaching its value at the upper critical dimension $d = d_c = d_f = 4$.

\begin{table}[h!]
\begin{center}
  \begin{tabular}{c|ccc}
    \hline \hline
    \hspace{5mm}$d$\hspace{5mm} & \hspace{5mm}$2$\hspace{5mm} & \hspace{5mm}$3$\hspace{5mm} & \hspace{5mm}$4$\hspace{5mm} \\ \hline
    $\beta$ & $1/8$ & $0.3265(3)$ & $1/2$ \\
    $\nu$ & $1$ & $0.6300(3)$ & $1/2$ \\
    $d_l$ & $15/8$ & $2.4817(5)$ & $3$ \\
    $\eta$ & $1/4$ & $0.0364(5)$ & $0$ \\
    $\nu$ & $1$ & $0.6300(3)$ & $1/2$ \\
    $d_f$ & $7/4$ & $2.9636(5)$ & $4$ \\
    $\zeta$ & $3/4$ & $0.9636(5)$ & $1$ \\
    \hline \hline
  \end{tabular}
  \caption{Values of critical exponents for the Ising universality class in 2, 3, and 4 dimensions. $d_l$ is from Eq.~\eqref{dl}, while $d_f$ and $\zeta$ are from Eq.~\eqref{dff}. Note the symmetric deviation for both the fractal dimension $d_f = d - \eta$ and the fractional derivative $\zeta = 1 - \eta$. For $d=3$, the results of Pelissetto and Vicari~\cite{Pelissetto02} are used. From reference \cite{Lima24}.}
  \label{Table1}  
\end{center}
\end{table}

\section{Conclusions}\label{sec:conclusion}

In this work, we discussed some major achievements in the formulation and evolution of the FDT. The breakthrough works of Einstein, Smoluchowski and Langevin established a simple and linear relation between the fluctuation and the dissipation in a physical system.

Moreover, within the approach the standard deviation appears as a very important source of information. 
Two basic examples of fluctuation phenomena - diffusion and random growth - were discussed along these lines.
Both of them have a powerful impact in physics and brought a large number of applications outside physical science. 
A few years after the origins of quantum mechanics, also the quantum fluctuations-dissipation theorems were born and with them a new physics emerged.

 The revolutionary use of the fluctuation-dissipation formalism in various domains of science continues and has gained new strength in the last two decades. { One recent example was the use of the growth fluctuation-dissipation theorem~\cite{GomesFilho21} to obtain the KPZ exponents~\cite{GomesFilho21b} and critical exponents in phase transition~\cite{Lima24}}. {Further generalizations of the FDT for a multivariate dynamical system close to a stationary state, as proposed by Zwanzig in a vector-matrix notation~\cite{Zwanzig01}, were recently investigated by considering various models of communication and information processing: chemical reaction networks~\cite{Chen_2022}, stochastic dynamics of geophysical systems~\cite{Weiss}, neural matrix models~\cite{Ewa2023,Enzo_2023} or performance and resilience of structurally perturbed graphs~\cite{Baggio}. In brief, the approach assumes a linearized dynamics in the form of multivariate Ornstein-Uhlenbeck process for which the FDT is recast as a continuous Lyapunov equation $\textbf{A}\textbf{C}+\textbf{C}\textbf{A}^T=2\textbf{D}$ for the dynamics described by coupled linear Langevin equations
\begin{equation}
\frac{d\vec{x}(t)}{dt}=-\textbf{A}\vec{x}(t)+\vec{\eta}(t).
\label{OU}
\end{equation}
Here $\textbf{A}$ is a friction matrix, $\vec{\eta}(t)$ is the noise vector, $\textbf{D}$ stands for a diffusion matrix $\left\langle\vec{\eta}(t')\vec{\eta}{^T}(t)\right\rangle=2\textbf{D}\delta(t'-t)$ and $\textbf{C}(t)=\left\langle\vec{x}(t)\vec{x}^T(t)\right\rangle$ is a correlation matrix. The reversibility of the process is guaranteed if $\textbf{A}\textbf{D}=\textbf{D}\textbf{A}^T$, otherwise the process attains a nonequilibrium stationary state \cite{Zwanzig01,Qian,Qian02,Luck,Landi_2013,Lucente_2025}. In this context one can further consider the entropy production rate $\Pi$ which provides a scalar measure of deviation of stationary state from full equilibrium. 
For a stationary diffusion process defined by Eq.~\eqref{OU} the entropy production rate takes the form of \cite{Landi_2013,Qian}
\begin{equation}
  \Pi =\left\langle\vec{x}^T(\textbf{D}^{-1}\textbf{A}-\textbf{C}^{-1})^T\textbf{D}(\textbf{D}^{-1}\textbf{A}-\textbf{C}^{-1})\vec{x}\right\rangle
  \label{EPR}
\end{equation}
where brackets stand for averaging over the stationary state probability density function. For reversible equilibrium dynamics Eq.~\eqref{EPR} yields zero and becomes positive whenever the symmetry of the matrix product $\textbf{A}\textbf{D}$ is violated~\cite{Zwanzig01,Landi_2013,Parrondo}.
The extent
to which a system infringed the FDT can provide insight into
its non-equilibrium activity; this issue remains of primary importance in debates about dominance of non-equilibrium processes in dynamics of living systems and networks of interconnected elements \cite{Gnesotto,Enzo_2023,Baggio,Kaluarachchi,Lucente_2025}. A critical aspect of linearized dynamics described by Eq.~\eqref{OU} is possible non-normality of the matrix $\textbf{A}$ ($\textbf{A}\textbf{A}^T\neq \textbf{A}^T\textbf{A}$). Usually linear/linearized systems are characterized by their spectrum. This approach is however not reliable for non-normal linear operators whose eigenvectors do not necessarily form an orthonormal basis. In the case of normal (symmetric) systems the largest eigenvalue of its spectrum is responsible for the long term dynamics and its inverse defines the characteristic relaxation time. In contrast, in non-normal systems more complex patterns may emerge, as the initial small perturbations around stationary states may undergo a transient evolution and be amplified \cite{nowak,Ewa2023_2}.
The non-normality of ${\bf A}$ has a dual impact: it entangles the dynamics of different modes through non-orthogonal eigenvectors, and it enhances the entropy production rate - a key marker of irreversible behavior \cite{Ewa2023_2}. These effects can be quantified by using the Chalker-Mehlig~\cite{Chalker} eigenvector overlap matrix $O_{ij}=\left\langle {L_i}|{L_j} \right\rangle \left\langle R_i|R_j\right\rangle$, which encapsulates the degree of non-orthogonality between left and right eigenvectors. The $O_{ij}$ matrix enters directly into modified expressions for correlation functions leading to a ``violation term'' $\Delta(\tau)={\bf AC}(\tau)-{\bf C}(\tau){\bf A}^T$ in the generalized FDT incorporating covariance ${\bf{C}}(\tau, t)=\left\langle \delta x(t+\tau)\delta x^T(t) \right\rangle=e^{{{\bf A}} \tau}{\bf {C}}(0, t)$ 
 that quantifies the departure from classical fluctuation-response symmetry and grows with the system's non-normality. Notably, even with identical eigenvalues, increasing eigenvector non-orthogonality amplifies this violation and the system’s entropy production $\Pi$, as captured by explicit trace formula~\cite{Ewa2023_2}, $\Pi\equiv - \frac{1}{2}\mathrm{Tr} {\bf D}^{-1}{\bf A}\Delta$. Importantly, these non-orthogonality-induced couplings produce strong transient amplifications, giving rise to collective effects such as synchronization and memory emergence in neural systems \cite{nowak,Ewa2023_2,Ewa2023}.

Recent contributions by Lucente \textit{et al.}~\cite{Lucente_2025,Lucente_2022} have further advanced the theoretical framework for analyzing fluctuation-dissipation relations in multivariate linear Langevin systems, particularly those exhibiting non-normal dynamics. These studies explore how the structure of the drift matrix, especially its non-normal components, impacts the entropy production rate and the form of the generalized FDT.

Abstraction of complex systems in terms of networks featuring nodes, edges and mutual interactions brings notion of non-normality of the adjacency matrix as representation of network topology. Symmetric interactions correspond to unidirected networks (any existing pair of edges between two nodes have the same weight in both forward and reverse directions) whereas asymmetric interactions correspond to directed networks with various strength of edges in different directions.
In consequence, linking the hierarchical, directed structure of networks with broken detailed balance, FDT violation and emergence of non-equilibrium dynamics become nowadays a fundamental task in unravelling structure, interaction strength and functionality of various complex systems in neuroscience, ecology, system biology and information processing \cite{Kaluarachchi}.}

The versions of quantum FDT, as derived by Callen and Welton~\cite{Callen51,Callen52} and generalized by Kubo~\cite{Kubo1957}, by referring to the linear response theory, were more recently extended to arbitrary quantum Markovian evolutions~\cite{Mehboudi} and its application to analysis of multipartite entanglement of complex quantum systems was investigated~\cite{Almeida2021,Zoller}.

In this context, recent advancements in quantum thermodynamics continue to yield exciting discoveries~\cite{Funo18,Sampaio18,Shibata20,Bull20,Hsiang20}, particularly regarding the quantum interpretation of FDT and its connection to fluctuations theorems~\cite{Xu_2023,Konopik}, as well as the recent extensions of FDT to active systems~\cite{Caprini18,Bechinger16, Caprini21,Burkholder19, Torrealba20}. It is quite certain that this line of research will be further explored in the next years to come.

\section*{Conflict of interest statement}

The authors declare that the research was conducted in the absence of any commercial or financial relationships that could be construed as a potential conflict of interest.

\section*{Author contributions}
All authors wrote sections of the manuscript. All authors contributed to manuscript revision, read, and approved the submitted version.

\section*{Acknowledgments}
We gratefully acknowledge Eli Barkai and Ismael Carrasco for their enlightening and stimulating discussions. This work was supported by the Conselho Nacional de Desenvolvimento Cient\'{i}fico e Tecnol\'{o}gico (CNPq), Grant No. CNPq-303119/2022-5, Funda\c{c}\~ao de Apoio a Pesquisa do Rio de Janeiro (FAPERJ), Grant No. E-26/203953/2022 (F.A.O.), E.G.N. acknowledges financial support by the Priority Research Area DigiWorld under the program Excellence Initiative - Research University at the Jagiellonian University in Kraków. 

\bibliographystyle{elsarticle-num}
\bibliography{reference1}

\end{document}